\documentclass[aps,prl,twocolumn,superscriptaddress,10pt]{revtex4-2}
\usepackage[dvips]{graphicx}
\usepackage{epsfig}
\usepackage{natbib}
\usepackage{amssymb}
\usepackage{url}
\usepackage{color}
\usepackage{times,fancyhdr}
\usepackage{amsmath}
\usepackage{amsfonts}
\usepackage{amsthm,amscd}
\usepackage{amsbsy}
\usepackage{latexsym}
\usepackage{bm}
\usepackage{layout}
\usepackage{makeidx}
\usepackage{epstopdf}	
\usepackage{color}
\usepackage{hyperref}
\usepackage[latin1]{inputenc}
\usepackage[T1]{fontenc}
\usepackage[ddmmyyyy]{datetime}
\def\be{\begin{eqnarray}}
\def\ee{\end{eqnarray}}
\def\veps{\varepsilon}
\def\der{{\rm d}}
\def\({\left(}
\def\){\right)}
\def\e{{\rm e}}
\def\d{{\rm d}}
\newdateformat{mydate}{\THEYEAR-\twodigit{\THEMONTH}-\twodigit{\THEDAY}}
\begin{document}
\title{Single-photon hot electron ionization of C$_{70}$}
\author{{\AA}ke Andersson}
\affiliation{Department of Physics, University of Gothenburg, 41296 Gothenburg, Sweden}
\author{Luca Schio}
\affiliation{IOM-CNR Tasc, SS-14, Km 163.5 Area Science Park, Basovizza, 34149
Trieste, Italy}
\author{Robert Richter}
\affiliation{Elettra - Sincrotrone Trieste, Area Science Park, 34149 Basovizza, 
Trieste, Italy}
\author{Michele Alagia}
\affiliation{IOM-CNR Tasc, SS-14, Km 163.5 Area Science Park, Basovizza, 34149
Trieste, Italy}
\author{Stefano Stranges}
\affiliation{IOM-CNR Tasc, SS-14, Km 163.5 Area Science Park, Basovizza, 34149
Trieste, Italy}
\affiliation{Dipartimento di Chimica e Tecnologie del Farmaco, Universit{\'a} 
Roma La Sapienza, Roma 00185, Italy Sapienza}
\author{Piero Ferrari}
\affiliation{Quantum Solid-State Physics, Department of Physics and Astronomy,
KU Leuven, 3001 Leuven, Belgium}
\author{Klavs Hansen}
\email{hansen@lzu.edu.cn,KlavsHansen@tju.edu.cn}
\affiliation{Lanzhou Center for Theoretical Physics, Key Laboratory of Theoretical Physics 
of Gansu Province and School of Physical Science and Technology, Lanzhou University, 
Lanzhou, Gansu 730000, China}
\affiliation{Center for Joint Quantum Studies and Department of Physics, School of Science, 
Tianjin University, 92 Weijin Road, Tianjin 300072, China}
\author{Vitali Zhaunerchyk}
\email{vitali.zhaunerchyk@physics.gu.se}
\affiliation{Department of Physics, University of Gothenburg, 41296 Gothenburg, 
Sweden}
\mydate
\begin{abstract}
Gas phase C$_{70}$ molecules have been ionized with single photons of energies
between 16 eV and 70 eV and the electron spectra measured with velocity map 
imaging in coincidence with the ions.
The doubly ionized and unfragmented species was present at photon energies of 22 
eV and up, and triply charged ions from 55 eV. 
The low kinetic energy parts of the spectra are explained with thermal emission of transient 
hot electrons. 
Deviations at high photon energies are used to determine a value for the initial 
electron equilibration time.
We propose a generally applicable mechanism, named Resonance Ionization Shadowing, 
for the creation of hot electrons by absorption of above-threshold energy photons. 
\end{abstract}
\maketitle

\section*{Introduction}\label{intro}

The large separation in time scales for electronic and vibrational motion of the nuclei 
opens the possibility of an intermediate phase of transiently hot electrons in 
molecules and clusters.
If present, this phase will exist between the time of the initial excitation of the 
electrons and the dissipation of the energy into vibrational motion. 
It tends to be manifested particularly clearly in finite systems, 
but has also been invoked in the description of the two-temperature model of 
solid surfaces exposed to short laser pulses \cite{AnisimovJETP1974}.

In gas phase context it was introduced as the explanation of the Penning ionization 
yields of C$_{60}$ and C$_{70}$ in Ref. \cite{WeberCP1998}.
Soon after it was observed also to be present in C$_{60}$ upon excitation with 
multiple low energy photons from laser pulses of duration around 100 fs 
\cite{CampbellPRL2000}.
Subsequently, the phenomenon successfully explained ionization of sodium clusters 
with short pulse laser light \cite{SchlipperAPA2001,MaierIJMS2006,MaierPRL2006}.
Following this development, it has been seen for a number of different systems
excited with short laser pulses, including C$_{70}$ \cite{KjellbergPRA2010} 
and a number of PAH (polycyclic aromatic) molecules \cite{KjellbergJCP2010}.

The dynamics of multi-electron excited states involved in the phenomenon 
has been considered theoretically
with different approaches in Refs. \cite{FlambaumPRA2002,
GribakinJPB2003,HeraudEPJD2021}, in addition to the more phenomenological 
models used to summarize the experimental results.
An integral part of this modeling when applied to molecules or clusters 
is the dissipation of the incoherent electronic 
excitation energy in the hot electron phase into the vibrational modes of the 
molecule. 
This coupling has been described in terms of a simple exponential decay of the 
excitation energy, involving a single parameter of dimension time, aptly named the 
coupling time. 
For some of the gas phase molecules studied, a proxy for this electron-phonon 
dissipation time has been measured by pump-probe experiments 
\cite{CampbellPPS2006,MaierPRL2006}.
In other cases it has been fitted from ion yield curves for different clusters 
\cite{HansenJCP2003}.
The values found range from a few hundred femtoseconds to a few picoseconds.
The fastest dissipation occurs for C$_{60}$, with a time constant of 240 fs 
\cite{HansenJCP2003}, and the slowest are the picosecond or longer times for 
sodium clusters \cite{MaierIJMS2006}. 
With reservation for the still limited number of systems studied at this point, the data 
point to a dependence of the coupling time that correlates positively with the average 
vibrational period, as given by the vibrational frequencies of C$_{60}$ 
\cite{SchettinoJPCA2001} and the bulk Debye temperature for sodium 
\cite{MaierIJMS2006}, although data from condensed phase nanoparticles 
show different trends \cite{MonginJPCM2019}. 
Those data pertain to much lower temperatures than relevant here, though.

The correlation seen in gas phase particles suggests a dissipation mechanism 
based on internal conversion, i.e. with similarities to the energy dissipation in 
molecules after absorption of single photons.
Experiments performed on thin films of C$_{70}$ have shown a very brief time 
window for equilibration, undetermined but below the pulse duration of 165 fs used in 
the experiments in Ref. \cite{WangCPL1996}.
Unfortunately it is not clear from these experiments if this time scale refers to the initial 
intra-electron equilibration or to the electron-phonon coupling time.

The initial electron equilibration in the creation of the hot electron phase has received 
much less attention experimentally than the final, dissipation stage.
It is clearly a subject of interest for the possibility of single-photon ionization of 
larger classes of molecules.
The observation of such single-photon hot electron ionization, already observed for 
C$_{60}$ \cite{HansenPRL2017}, opens the possibility for studies of the 
mechanisms of absorption and initial dissipation of the energy.
In addition to the general relevance for delineating the boundaries of the mechanism,
single-photon excitation is also of interest in astrophysical context because 
molecular ions play an important role in the interstellar chemistry \cite{Herbst_2005},
and in particular for fullerenes because they have been identified in the interstellar medium.
The experiments reported here on C$_{70}$ were motivated by the above 
questions.
As an aside we mention that single-photon processes come with the additional 
and very attractive feature that they eliminate the uncertainty in energy that 
accompanies multi-photon processes previously used for studies of the subject. 
The experiments will also allow a test of the interpretation of the previous results on 
C$_{60}$ in Ref. \cite{HansenPRL2017}, using a molecule with almost equally 
well characterized and similar but still different properties. 

The clearest experimental signature for these purposes remains the emission of electrons 
that are thermalized to the very high energies which characterize the hot electron 
phase.
The emission of electrons that can be unambiguously assigned as hot electrons 
occurs between the initial excitation and the dissipation of energy into the 
vibrational motion.
These distributions are unique to hot electron emission, and have the added 
experimental convenience that the spectra do not need to be measured 
time-resolved.
However, ionization may also occur both before and after the creation of the hot 
electron phase.
Either by direct ionization, which may remove enough energy by the departing 
electron to preempt the creation of the hot electron phase, or by thermionic 
emission after dissipation of the energy into the predominantly vibrational 
excitations of the equilibrium state.

The form of the thermal electron spectra is shaped by a number of factors
\cite{AndersenJPB2002}.
One is the product of the emitted electrons' phase space and a flux factor in the 
form of the speed of the emitted electrons.
These combine to give a factor proportional to the kinetic energy of the channel.
A second factor is the cross section for the inverse (attachment) reaction.
The third and last factor is the ratio of the level densities of the product and emitting 
molecules \cite{HansenJCP2003}.
These factors enter the expression for the electron kinetic energy-resolved 
rate constants, which is identical to the one for the usual thermionic emission 
apart from the different level densities that describe the emitting systems in 
the two situations. 
The phase space and the speed factors combine to give the electron kinetic energy 
to the power one.
For neutral or positively charged emitters the cross section of the inverse 
process of absorption is basically that of a Coulomb potential. 
In a classical calculation, which will be used here, it is proportional to the 
reciprocal of the electron energy, plus a constant
(see ref. \cite{HansenJCP2003} for details).
The ratio of level densities acts as an effective Boltzmann factor.
The net result is that for neutral and positively charged emitters, the energy 
distributions calculated under these assumptions resemble Boltzmann factors 
with the effective temperatures given by the product microcanonical electron 
temperature, as discussed in \cite{AndersenJCP2001}.
For more information on the derivation of the expression, please see Ref. \cite{book}.
The very good consistency of several different experimentally measured quantities
with the predictions derived from this description reported in \cite{HansenJCP2003}
constitute a strong support of the modeling.

In addition to the Boltzmann-like shape of the spectrum, there are several 
other features that makes it distinct from the spectra originating either from 
direct ionization or from thermal emission from completely equilibrated 
molecules, known as thermionic emission.
A necessary feature of the spectra is that the velocity distributions of the emitted 
electrons must be spherically symmetric. 
This is a property shared with electrons emitted into single particle s-states, 
and for a single-photon excitation this could explain this symmetry, albeit not 
the Boltzmann shape.
However, the energies of such electrons and indeed all electrons emitted from 
single-particle states move in parallel with the photon energy and will therefore have 
a different photon energy dependence than the hot electron spectra.
Measurements at a few different photon energies are therefore sufficient to 
distinguish an origin of the relevant low energy part of the spectra as thermal 
or as emitted in a direct process.

A third possible origin of electrons, besides the hot electron emission and the 
direct ionization, is a regular thermionic process.
There are two important differences between this type of process and hot 
electron emission.
One is the effective temperature of the Boltzmann distribution.
A standard thermionic emission process comes with an internal energy which 
renders the effective (microcanonical) temperature much lower than the hot 
electron emission.
For fullerenes, for example, the thermionic emission temperature has been 
fitted to values around 3500 K from electron spectra measured with the  
velocity map imaging (VMI) technique also used in this work 
\cite{LepinePRA2004}. 
Although this is a very high temperature in many connections, the very fast 
emission required for the hot electron system requires much higher temperatures, 
on the order of 1 eV (= 11605 K) and higher \cite{HansenJCP2003}.
The fitted temperature for the one photon hot electron ionization of C$_{60}$
reported in  \cite{HansenPRL2017} reached 1.6 eV, for example.

The other difference to hot electron ionization is the much longer time scale 
on which thermionic emission can be observed. 
Hot electron emission is limited to picosecond or sub-picosecond time scales.
Thermionic emission, in contrast, will, for low excitation energies, extend 
to time scales that under some conditions can be detected as a several 
microsecond long tail on the mass peak in time-of-flight mass spectra 
\cite{WalderIJMPB1992}
As a secondary signature, thermionic emission from neutral and cationic 
fullerenes is usually observed together with a substantial amount of fragmentation.
Their absence here is only corroborative for the absence of thermionic emission, 
though.

For the doubly ionized species observed in the experiments here, two other 
possible channels should be considered.
One is the direct double electron ionization. 
The electrons associated with prompt double ionization are characterized 
by a U-shaped electron kinetic energy distribution \cite{AnderssonSR2019}.
The steepness of these distributions depend on the relation between photon 
energy and the double ionization potential values.

Another possible channel is the emission of a second electron by regular 
thermionic emission.
This process would occur after the excitation energy has been dissipated 
into the predominantly vibrationally excited equilibrium state.
However, this is ruled out for two reasons. 
One is that the competing C$_2$ loss channel would dominate over thermionic 
emission by a large factor. 
The second is that delayed emission is absent in the time-of-flight mass 
spectra for the double charged species.

In summary, the nature of the emission process is very well established, and 
displays the primary characteristic experimental features listed above. 
A number of further derived features can be found in \cite{HansenJCP2003,
LassessonEPJD2005}, where experiments on fullerenes and endohedral 
fullerenes exposed to short pulses of photons with energies below the 
ionization energy are described.

Apart from the initial Penning ionization study on C$_{60}$ and C$_{70}$ 
\cite{WeberCP1998} and the one photon hot electron ionization study in Ref. 
\cite{HansenPRL2017}, all of the experimental studies mentioned have been 
performed with short laser pulses of sub-threshold photon energies. 
Indeed, until recently, studies of the hot electron phenomenon with photon 
excitations has been limited to excitation with photon energies below the ionization 
threshold and, therefore, to excitation energies provided by absorption of several 
and often a large number of photons.
It is not a priori clear whether a single photon can cause the creation of the hot 
electron phase.
Clearly, the competing direct ionization (spectroscopic) channel also appears 
prominently.
For a recent such relatively low photon energy study of C$_{60}$, see 
\cite{HrodmarssonPCCP2020}.
However, in \cite{HansenPRL2017} it was shown that the absorption of single 
high energy photons by C$_{60}$ can indeed give rise to hot electron emission, 
in parallel with the direct ionization of standard spectroscopic nature.
The experimental signatures used to establish this were the ones associated 
with hot electron emission known from the multi-photon experiments and 
listed above, supplemented by the appearance energies of the fragmented ion
at photon energies far above the ionization energy. 
The present experiment aimed to explore if and how these results applied 
also to C$_{70}$.

\section{Experimental and theoretical procedures and results}

The experiments were performed at the GasPhase beam-line at the synchrotron 
ring Elettra.
The procedures were similar to those used for C$_{60}$ \cite{HansenPRL2017}
and only a brief description will be given here. 
For further specifics of the beam line the reader is referred to 
\cite{BlythESRP1999,OkeeffeRSI2011}.
The fullerene material was acquired from Sigma Aldrich, with a purity of 
98 \%.
The sample was heated in situ for five days above 200$\,^{\circ}$C to out-gas 
solvents and other volatile contaminants. 
During measurements the molecules were sublimed from an oven with a temperature 
of initially 430 $\,^{\circ}$C, slowly increasing to 470 $\,^{\circ}$C at the end of the 
run, in order to keep the evaporation rate constant.
The temperature was measured by a thermocouple attached to the oven.
The linearly polarized light was filtered by standard filters at the different
wavelengths as needed. 

The electron spectra were recorded on single-count basis with a VMI 
spectrometer equipped with a dual delay line position sensitive detector and 
analyzed off-line.
The coincidence of electrons and ions was extracted offline from the recorded
time of flight of the ions with a electron detection defining zero time.
The detector only allowed detection of a single electron per event.
To reduce the amount of false coincidences, the light intensity was reduced 
to electron count rates of 11--18 kHz and ion count rates of 2--8 kHz.
As the detection efficiency does not depend on the origin of an electron, 
the spectra of molecules with a specific charge state are equal weight average 
spectra of all emitting charge states leading to the final state. 

\begin{figure}
\centering
\includegraphics[width=0.45\textwidth,angle=0]{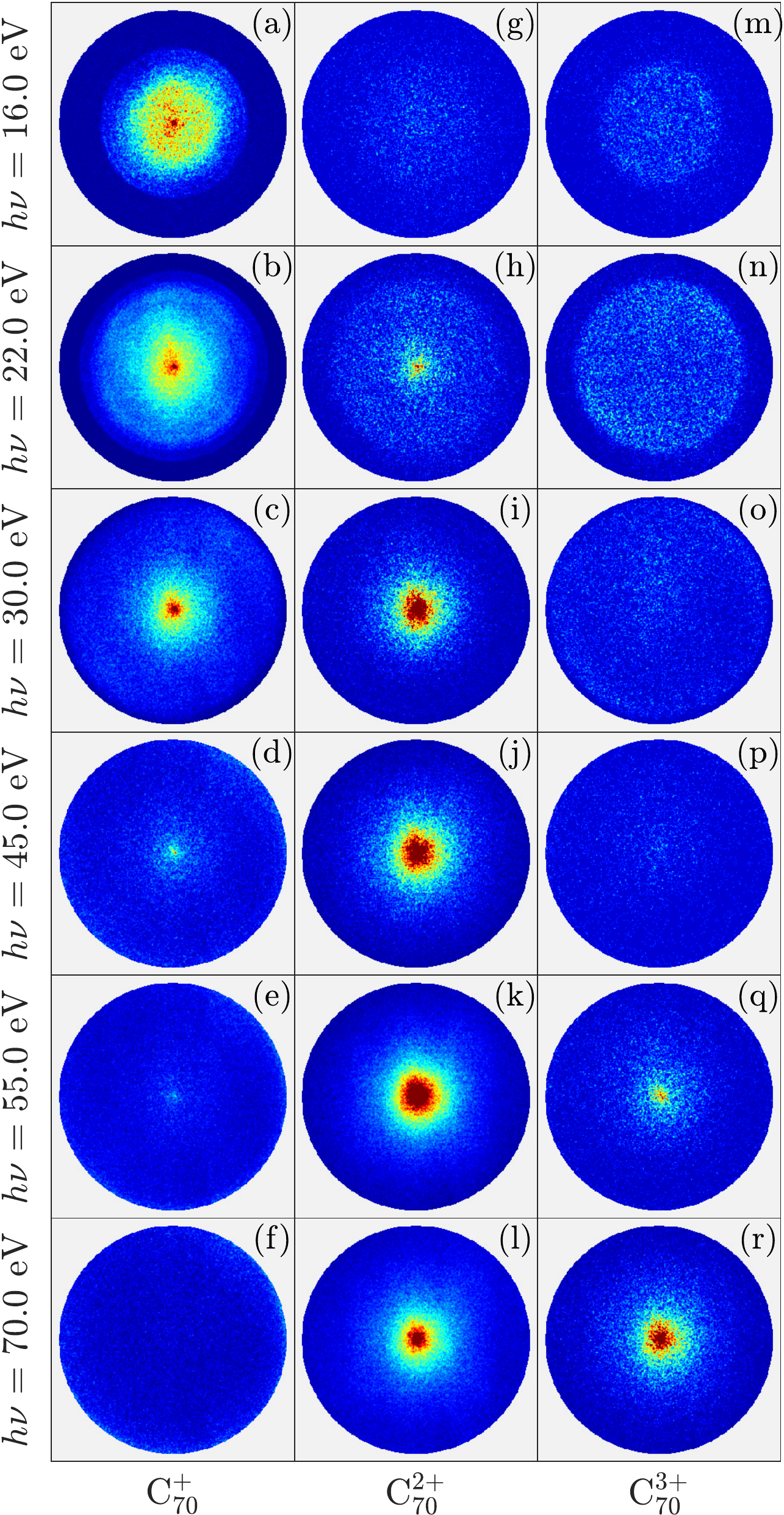}
\caption{\label{ManyVMI}
The raw spectra of, from left to right,  C$_{70}^+$, C$_{70}^{2+}$, and 
C$_{70}^{3+}$.
The narrowing of the intensity into a low momentum peak in the C$_{70}^+$
column results from the transformation of the ionization process from direct to 
hot electron emission.
As seen, the second ionized species is first visible at 22 eV and 
the triply charged at 55 eV. 
The strong intensity of the highly charged species contrasts with the result for 
C$_{60}$, for which the high photon energy spectra are dominated by the 
fragments.}
\end{figure}

Spectra were recorded for the photon energies 16, 20, 22, 23.8, 26, 28, 30, 35, 
40, 45, 50, 55, 60, 65, and 70 eV, always with the polarization parallel to the VMI 
detector plane. 
No tails on the mass spectra that would indicate a thermionic emission from a 
completely equilibrated system were observed in this experiment.
Likewise, the substantial fragmentation that accompany thermionic emission
for fullerenes was absent in the C$_{70}$ mass spectra recorded in this work,
as in our previous work on C$_{60}$ \cite{HansenPRL2017}.
The raw data spectra for the three lowest measured charge states of the unfragmented 
molecule at several selected photon energies are shown  in Fig. \ref{ManyVMI}.

\begin{figure}
\centering
\includegraphics[width=0.45\textwidth,angle=0]{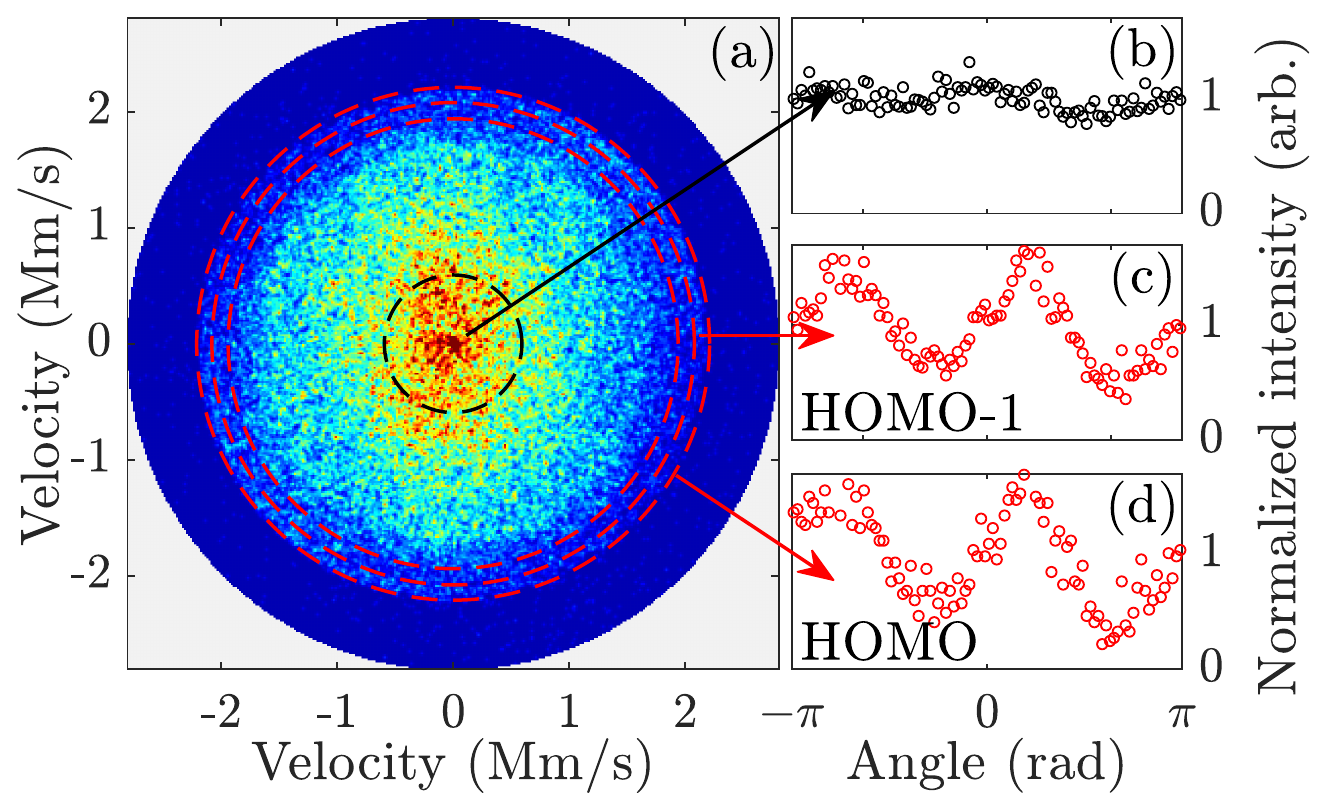}
\caption{\label{pizzaVMI_20}
A measured spectrum for C$_{70}$ at photon energy 20 eV. 
Frame (a) shows electron intensities across the detector surface.
Frame (b) shows the angular resolved intensities for the low energy part of the 
spectrum, and two high energy electron parts defined by the red circles are 
shown in frames (c,d).
The flat distribution in frame (b) is consistent with a spherically symmetric 
momentum distribution, in contrast to the direct ionization electrons in frames 
(c,d).}
\end{figure}

The measured VMI spectra are the momentum distributions of the emitted 
electrons projected on the detector plane.
On the VMI detector surface the required spherical symmetry of the hot electrons 
corresponds to a circular symmetry.
Figure \ref{pizzaVMI_20} shows the angular symmetry at the low energy 
electrons and the contrast to the asymmetry for higher energy electrons for
a spectrum recorded after exposure to 16 eV photons. 
At low photon energies, the spectrum contains a wide base with structures 
that can be identified as features of direct ionization and hence of spectroscopic 
nature. 
The intensity peaking at zero kinetic energy, together with the appearance of 
the circular symmetry of these parts of the spectra, indicates the emergence of 
the hot electron spectra.
Indeed, for all photon energies the central, lowest energy part of the spectra 
showed no sign of a correlation of the intensity with the direction of the light 
polarization, indicative of the required symmetrical distributions. 

With the chosen light polarization the complete 3D distributions are obtained 
from the VMI spectra by deconvolution.
The deconvolution was done with the inverse Abel transform as implemented 
in the MEVIR software \cite{DickPCCP2014}.
Deconvolution of the spectra requires that the entire spectrum is projected onto
the VMI detector surface.
The highest electron energy for which this is guaranteed was 23 eV for the 
VMI voltages used in the experiment.
This limits the photon energies to below 23 eV $+E_{\rm i,1}$ for the singly 
ionized species, with $E_{\rm i,1}$ the first ionization energy.
The value of $E_{\rm  i,1} =7.4$ eV was measured in Ref. \cite{BoltalinaJACS2000}, 
making this limit equal to 30.4 eV. 
A conservative safety margin on the masking reduces the highest photon energy 
to 26 eV for the singly charged species.

\begin{figure}[ht]
\centering
\includegraphics[width=0.45\textwidth,angle=0]{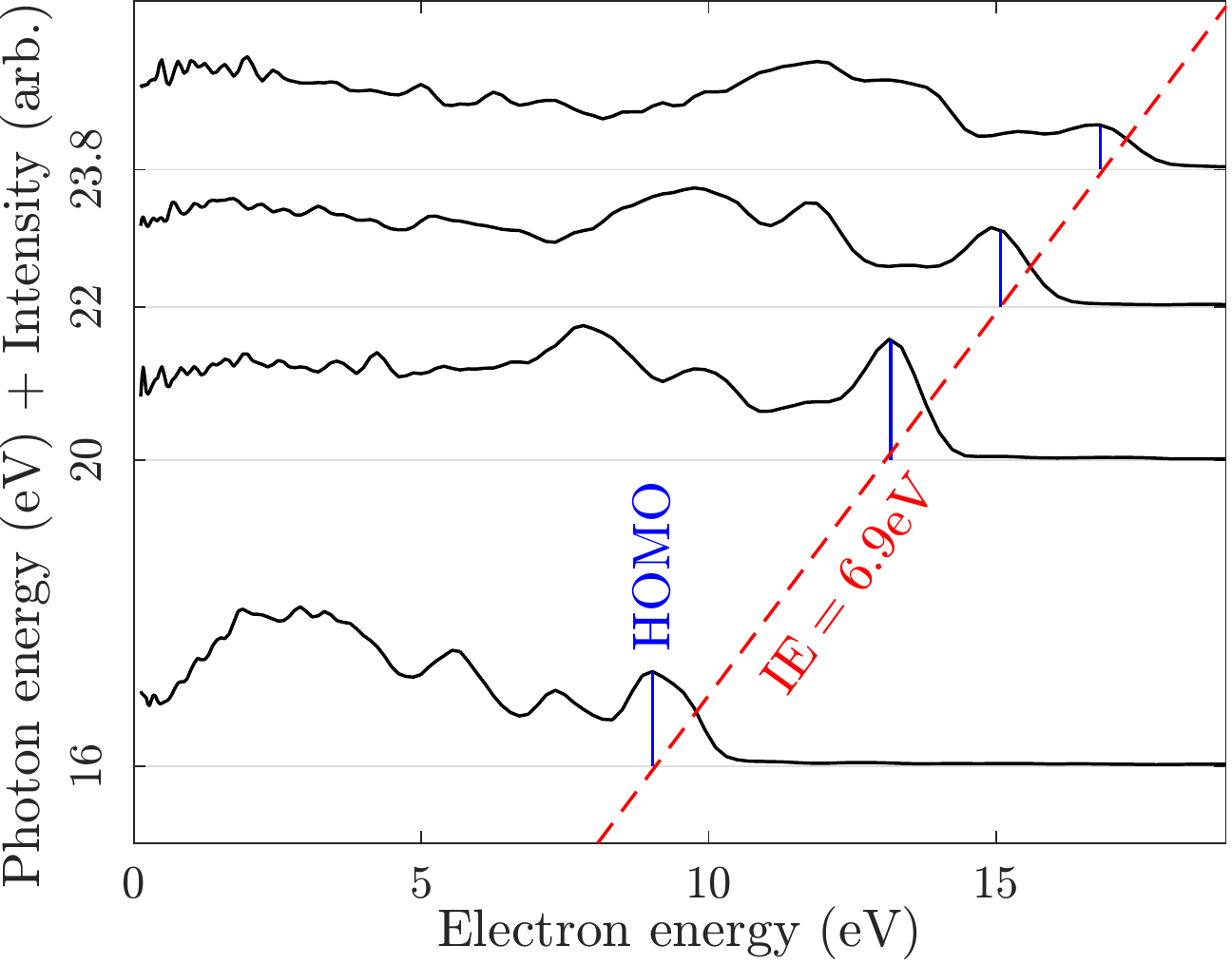}
\caption{\label{IE}The determination of the first ionization energy of C$_{70}$ from 
the measured direct ionization spectra.}
\end{figure}

As a check of the procedure, the value of the ionization energy can be inferred by 
tracing the position of the highest occupied molecular orbital (HOMO) level 
as a function of the photon energy.
The four photon energies from 16 eV to 23.8 eV can be used for that purpose.
Figure \ref{IE} shows the trace used to determine $E_{\rm i,1}$ on the VMI spectra 
deconvoluted with the procedure which is explained in more detail below.
The value from this determination is $6.9\pm 0.5$ eV, where the uncertainty is 
mainly due to the width of the peaks, i.e. consistent with the value from Ref. 
\cite{BoltalinaJACS2000}.

The electron detector can only assign a position and hence a transverse momentum 
to an event when it is hit by a single electron.
As the detection efficiency is less than unity, it is therefore nevertheless still 
possible to detect spectra from double and triple ionization events.
In these cases the spectra are sums of two spectra (for double ionization) or 
three spectra (for triple ionization) with equal weights.
Since the detection limit is 21 eV and the sum of the two lowest ionization 
energies is 18.84 eV \cite{StegerCPL1992}, all the electrons originating from 
double ionization events at photon energies below 40 eV are within detection 
range.
For comparison, the value calculated with density functional theory for the 
second ionization energy is 10.3 eV (see below for the method used). 
With the calculated single ionization value of 7.3 eV this is in reasonable albeit 
not perfect agreement with the measured value.
We have used the most conservative, experimental value.
The appearance of the doubly charged ions at the photon energy $h\nu = 22$ eV
is higher than the literature and theoretical values, as expected, and is consistent 
with the interpretation of the origin of the emitted electrons.
In summary, inversion was performed for photon energies up to 26 eV for singly 
charged species and up to 40 eV for the doubly charged molecules, and for the 
latter only for single-electron events.

\begin{figure}
\centering
\includegraphics[width=0.45\textwidth,angle=0]{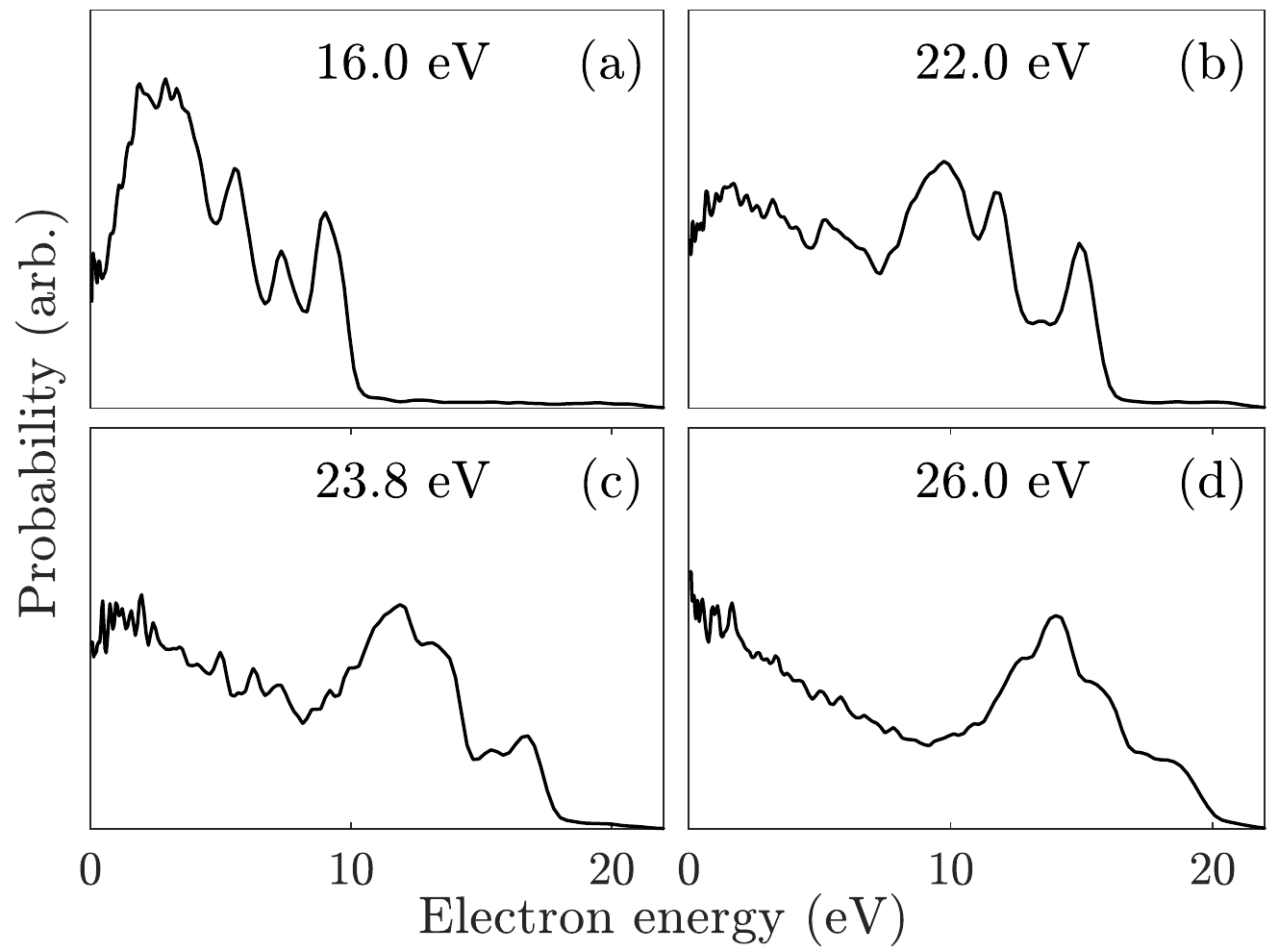}
\caption{\label{EES_840_2x2}
The angular integrated electron spectra measured in coincidence with 
C$_{70}^+$ for the a series of photon energies.
The $h\nu$ spectrum is the highest measured spectrum that can be deconvoluted 
for singly ionized species. 
The direct ionization intensity, which for the $h\nu=26$ eV spectrum is found 
between 10 and 20 eV, and the hot electron 
intensity, up to 10 eV, can be compared directly and are seen to be very 
similar at this photon energy.}
\end{figure}

\begin{figure}
\centering
\includegraphics[width=0.5\textwidth,angle=0]{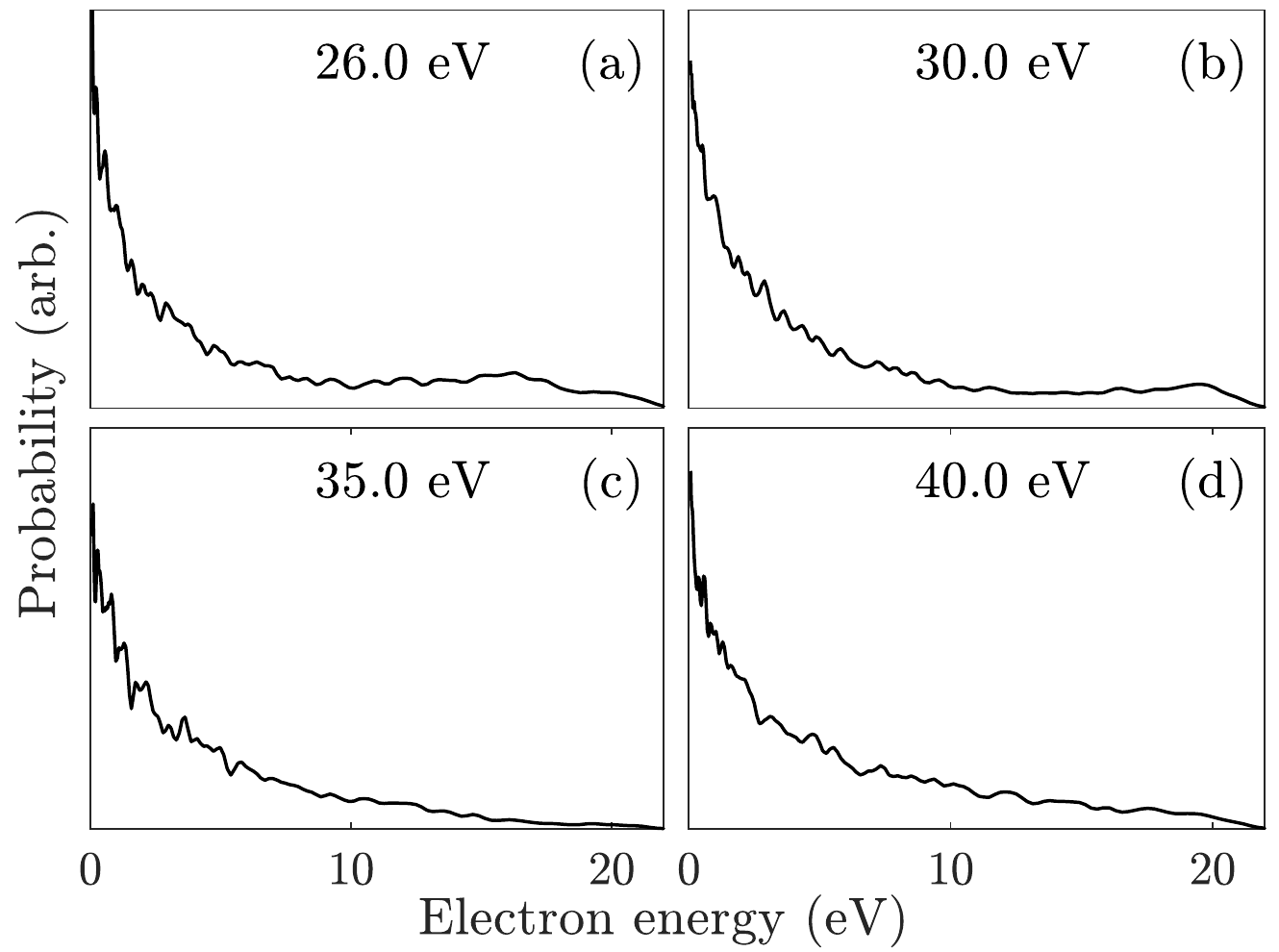}
\caption{\label{EES_420}
The angular integrated electron spectra measured in coincidence with 
C$_{70}^{2+}$ for the photon energies indicated in the frames.}
\end{figure}

The spectroscopic nature of the high energy electrons is more apparent after 
deconvolution of the spectra. 
Figure \ref{EES_840_2x2}, which shows deconvoluted spectra, demonstrates 
how this picture develops with increasing photon energy, in particular how the low 
energy electrons become increasingly intense.
Figure \ref{EES_420} shows the spectra of single electrons detected 
in coincidence with doubly charged ions.
The potential competing process of the low energy electrons by direct double ionization
is not seen in the spectra in Fig. \ref{EES_420}.
These distributions would have a U-shape and the high energy end of such a spectrum 
would be present at the high kinetic energies, which is clearly not the case. 
We can therefore rule out this channel as a significant contribution also to the low 
energy part of the spectra.

Quantum chemical calculations of total and individual level energies were 
performed with density functional theory (DFT) on C$_{70}$ using the 
ORCA 5.1 software package \cite{neese2020orca}. 
For this, the PBE functional \cite{hammer1999improved} was used with 
the Def2-TZVPP basis \cite{weigend2005balanced}, and included dispersion 
corrections via the D3BJ approximation \cite{grimme2011effect}. 
The geometries of both molecules were optimized for the charge states 0, 1, 2
and all the electrons were included in the calculation.
Moreover, vibrational frequencies were computed, confirming that structures 
represent true minima on the potential energy surface. 
In addition, single-point calculations of C$_{60}^{q+1}$ and C$_{70}^{q+1}$ 
on the optimized geometries of C$_{60}^{q}$ and C$_{70}^{q}$ were 
performed for $q=0,1,2$, in order to calculate vertical first and second ionization 
energies.

Although the use of DFT to calculate the ionization energies and vibrational 
frequencies are well controlled, it is relevant to add a remark about the use of
the single particle pseudo-states to calculate level densities. 
These do not in principle give the single particle states, and neither do they 
guarantee that the single particle approximation can be used.
However, it is possible to compare level densities calculated with those states 
with the experimentally determined level density.
The comparison was made in \cite{HansenJCP2003} where the two were found to 
agree very well.
The only fit parameter used in that comparison was a single multiplicative constant
which provided the absolute magnitude which can not be extracted from the 
experiments.
We will therefore use the same procedure here to calculate the level densities.

\section{Analysis and discussion}

Before a quantitative analysis of the deconvoluted spectra is presented, 
it is of interest to 
consider the raw data plot in Fig.\ref{ManyVMI} in some detail.
Important features of the processes here and for C$_{60}$ (see Fig. 1 of
\cite{ReinkosterJPB2004} and Fig. 5 of \cite{HansenPRL2017}) are that\\
\textit{i)} the dominant open decay channel for the singly ionized  C$_{70}$,
shown in the second and third columns, is further electron emission 
producing the higher charge states of the molecules.
For C$_{60}^+$ fragmentation is somewhat more pronounced.\\
\textit{ii)} 
intensities for C$_{70}^{2+}$ (and for C) appear at lower photon energy than 
the fragments of C$_{60}^+$, and \\
\textit{iii)} the difference in appearance energies of
the triply and doubly ionized charge states of both C$_{60}$ and C$_{70}$ 
is much larger than the corresponding 
difference between the appearance energies of C$_{58}^+$ and C$_{56}^+$
from C$_{60}$.

Concerning \textit{i)}, the tendency to ionize twice instead of causing 
fragmentation was already reported in \cite{MitsukeJPCA2007}.
A similar effect has been seen in naphthalene \cite{ReitsmaJPCA2019}, where 
the relative intensities of doubly ionized species relative to singly charged 
fragments increases when photon energies are changed from 20.4 eV to 29.8 eV,
in parallel with a strong suppression of fragmentation processes at the higher 
energies.

The explanation of \textit{ii)} is that the second ionization occurs from the 
hot electron ensemble, whereas the fragmentation of C$_{60}^+$ ions occurs 
from the completely vibrationally thermalized ion.
The difference in the heat capacity of the two emitting systems accounts for the 
main part of this difference in appearance energies.
These aspects have already been analyzed in detail in \cite{HansenJCP2003} and 
\cite{HansenPRL2017}, where more quantitative details can be found.
We note that also this observation is consistent with the hot electron ionization
mechanism.

The reason for the behavior in point \textit{iii)} is also the different nature of the 
decays of the two molecules.
Addressing this question requires quantitative considerations of the appearance 
energies.
For the C$_{60}$ decay, the difference in the appearance energies of 
C$_{58}^+$ and C$_{56}^+$ is given mainly by the C$_2$ dissociation energy 
of C$_{58}^+$. 
This is seen with the following simplified but still reasonably accurate calculation. 
The appearance energy for fragment $m$ in a decay chain can be calculated as 
the photon energy which is the sum of the energies consumed in the previous decays 
plus the thermal energy needed for the $m$'th decay.
With an Arrhenius expression for the fragmentation rate constant we have 
\be
k(E) \sim \omega \exp\(-D/T(E)\),
\ee
with $T$ being the effective microcanonical temperature at that energy $E$, and $D$
the evaporative activation energy.
A linear relation between the (microcanonical) vibrational temperature $T$ 
and the excitation energy $E$ is assumed ($E_m^{\rm o} $ is the energy offset in this 
curve, and $k_{\rm B}$ is set to unity):
\be
E = C_kT - E_m^{\rm o}.
\ee  
With $G$ defined as $\ln{\omega t}$ and $t$ being the time of acceleration for the ion 
time-of-flight, this gives 
\be
h\nu_{{\rm appear},m} \approx D_m \frac{C_m}{G} + E_m^{\rm o} + 
\sum_{j=0}^{m-1}D_j - E_{\rm source}.
\ee
The last terms in the equation accounts for the energy consumption in the prior 
decays and the initial energy of the molecule.
The small amounts of energy carried away by the C$_2$ fragments are ignored in the expression. 
When the contribution $D_m C_m/G +E_m^{\rm o}$ is approximately 
independent of $m$, the difference in the $m$'th and the $m-1$'th appearance 
energies is
\be
h\nu_{{\rm appear},m} -h\nu_{{\rm appear},m-1} \approx D_{m-1}.
\ee

This approximate identity of the sequential differences between 
appearance energies hinges on the similarity of the emission activation 
energies and the constancy of the heat capacity.
These are expected to hold to a decent approximation for the C$_2$ loss
activation energy and for the vibrational thermal properties.
For the hot electron emission processes seen for C$_{70}$, neither of these 
similarities will hold.
The emission activation energies are the ionization energies, and their values
increase with the charge state.
Also the heat capacities vary with energy.

The estimate for the electron emission appearance energies goes as follows:
The lowest effective temperature where hot electron emissions occur is 
determined by the combination of the electron-vibrational cooling time, 
which we will set to the C$_{60}$ value of $\tau = 240$ fs 
\cite{HansenJCP2003} and the electron emission rate constant by the relation 
\be
\label{k-limit}
k(E) = 1/\tau.
\ee
for smaller rate constants, dissipation into vibrational motion quenches the emission.
The emission rate constant for electrons is also written as an Arrhenius expression
where the activation energy is the ionization energy $E_{\rm i}$.
The frequency factor is denoted by $\omega_{\rm e}$.
Although the value of $\omega_{\rm e}$ depends on the charge state, the 
dependence is minor and beyond the precision here, and the factor will therefore be 
set to the neutral molecule value.
To find the temperature we use the caloric curve for a Fermi gas, 
\be
E = \frac{1}{2}\alpha T^2.
\ee
The initial electronic energy from the source can be set to zero.

The photon energy at which the second ionized species appears can then be 
calculated with the same logic as for the unimolecular decays, i.e. adding the 
consumed energies of the previous decays to the excitation energy calculated by
Eq. \ref{k-limit}.
The result is 
\be
\label{hnu2}
h\nu_2 =
\frac{\alpha}{2 (\ln \omega_{\rm e} \tau)^2} E_{\rm i,2}^2 +E_{\rm i,1} + \langle \veps_{1,1}\rangle,
\ee
where $\langle \veps_{1,1}\rangle$ is the average electron energy in the first ionization.
By the same argument the triply ionized species appear at the photon energy
\be
h\nu_3 = 
\frac{\alpha}{2 (\ln \omega_{\rm e} \tau)^2} E_{\rm i,3}^2 + E_{\rm i,1} + E_{\rm i,2} 
+\langle \veps_{2,1}\rangle + \langle \veps_{2,2}\rangle,
\ee
where $\langle \veps_{2,1}\rangle$ and $\langle \veps_{2,2}\rangle$ are the electron 
energies of the first and second emitted electron in this process. 
These energies are larger than the counterpart for C$_2$ emission and can not be 
ignored in the analysis for electron emission.
The emission of the first electron occurs at different energies for the two processes with 
the different final charge states, and $\langle \veps_{1,1} \rangle$ is therefore different 
from (smaller than) $\langle \veps_{2,1}\rangle$.
With the value 26 eV for $h\nu_2$ (see Fig. \ref{ManyVMI}), the first 
ionization energy $E_{\rm i,1} =7.4$ eV, and the second $E_{\rm i,2} = 11.4$ eV, the 
coefficient in Eq.\ref{hnu2} becomes 
\be
\label{Ehnu2appear}
\frac{\alpha}{2 (\ln \omega_{\rm e} \tau)^2} = \frac{h\nu_2 - 
E_{\rm i,1} - \langle \veps_{1,1}\rangle}{E_{\rm i,2}^2} = 0.13 ~{\rm eV}^{-1},
\ee
when we use the value $\langle \veps_{1,1}\rangle = 2$ eV.
This result be compared below with the theoretical value derived from the rate constant
after that calculation has been made.

Using the close similarity of the first two ionization energies to those of C$_{60}$,
7.6 eV and 11.4 eV, respectively, we adopt the third ionization energy of C$_{60}$, 
$E_{\rm i,3} =16.6$ eV, for C$_{70}$.
This predicts an appearance photon energy of the third ionized molecule of
\be
h\nu_3 = (18.8 + 5)~{\rm eV} + 0.13 ~{\rm eV}^{-1} (16.6~{\rm eV})^2 
= 60 ~{\rm eV},
\ee
where by inspection of the measured spectra shown in Fig. \ref{EES_420} we 
estimated the sum $\langle \veps_{2,1}\rangle + \langle \veps_{2,2}\rangle$
to be 5 eV.
The estimated uncertainty on $h\nu_3$ is 9 eV.
The experimental value of this cross-over photon energy is more uncertain than 
$h\nu_2$, but the above calculated value is within the range of the possible 
experimental values that lie between 45 eV and 60 eV.

For the above analysis a description in terms of a Fermi gas is sufficient, but for a more 
precise description and an assessment of the value of $\alpha$, a more accurate 
calculation of the thermal properties of the hot electrons is required.
The relevant thermal properties are the level densities, or density of states and 
the rate constants.
They are calculated with the method given in the appendix of Ref. 
\cite{HansenJCP2003}.
The input data are the energy levels from the DFT calculation of the energies. 
As the temperature is the microcanonical version we use the value derived 
from the level density $\rho$ \cite{AndersenJCP2001}
\be
\label{mictemp}
\frac{\d \ln \rho( E )}{\d E} = T^{-1}.
\ee 
As $k_{\rm B}$ is set to unity, temperatures are therefore given in eV.
This calculation for the singly charged molecule essentially confirms the 
Fermi gas Ansatz, albeit with an offset in the temperature.
The fitted form is 
\be
\label{caloric}
E = \frac{1}{2}\alpha' \(T^2 - T_0^2\),
\ee
with $\alpha' = 46$ eV$^{-1}$ and $T_0^2 = 0.3$ eV$^2$.
These values pertain to the singly charged molecule but the values for the other
charge states are similar.
As a side remark we note that the offset in the caloric curve is analogous to the 
similar and well documented offset that appears in the caloric curve of quantized 
vibrational motion. 
The interpretation of the offset here is different and the value can not be ascribed 
to a zero point motion as for the vibrations.
The offset in temperature prevents a comparison of the $\alpha'$ fitted here 
and the $\alpha$ calculated from Eq. \ref{Ehnu2appear} with the rate constant 
in Eq. \ref{weissk}. 
This comparison will be made below.

\begin{figure}
\centering
\hspace*{-1.5cm}
\includegraphics[width=0.45\textwidth,angle=90]{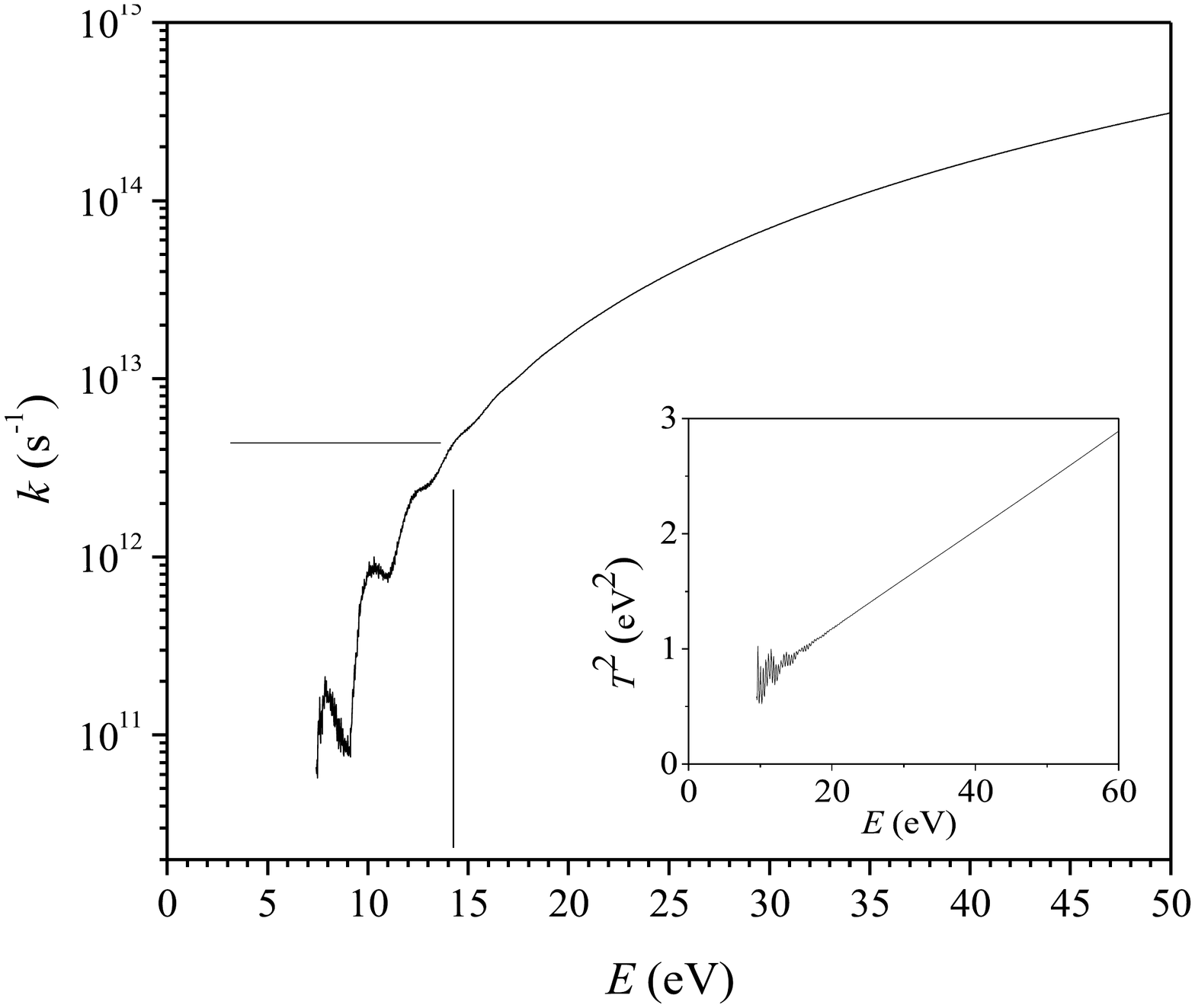}
\vspace{-1cm}
\caption{\label{k2}
The rate constant for emission of the second electron.
The reciprocal coupling time and the corresponding lower limit of the energy 
of an emitting molecule are indicated by the horizontal and vertical lines.
The inset shows the square of the microcanonical electron temperature vs. 
excitation energy.
This is clearly very well represented by Eq. \ref{caloric}.}
\end{figure}

The kinetic energy-resolved electron emission rate constant is given by the 
expression \cite{HansenJCP2003}
\be
\label{weissk}
k^{(q)}(E,\veps)\der \veps =
\frac{2m_{\rm e}\sigma (\veps )}{\pi ^{2}\hbar ^{3}}
~\veps ~
\frac{\rho^{(q+1)}(E-E_{\rm i, q} -\veps)}{\rho^{(q)}(E)}
\der \veps.
\ee
Here $\varepsilon $ is the kinetic energy of the electron, $m_{e}$ is the mass of 
the electron, $\rho^{(q)}(E)$ is the level density of charge state $q$ at energy $E$, 
and $E_{\rm i,q}$ is the ionization energy.
The factor of two in Eq.\ref{weissk} is the spin degeneracy of the electron and
$\sigma (\varepsilon )$ is the capture cross section for an electron in the Coulomb 
potential of the decay product.
To find the relevant rate constant, the kinetic energy is integrated over:
\be
\label{weissk3}
k^{(q)}(E) \equiv \int_0^E k^{(q)}(E,\veps) \der \veps.
\ee
The numerically integrated function is shown in Fig. \ref{k2}.

With Eq. \ref{weissk} we can describe the kinetic energy distributions with 
the function
\be
\label{weissk2}
P(\veps;E,q) \der \veps =
\sigma (\veps )~\veps ~\rho^{(q+1)}(E -\veps) \der \veps.
\ee
For the capture cross section the classical values used are
\be
\sigma(\veps) = \pi r_0^2 \(1- \frac{V(r_0)}{\veps}\),
\ee
and
\be
V(r_0) = -\frac{(q+1) e^2}{4\pi\veps_0 r_0} 
= -(q+1) 3.0 ~{\rm eV}.
\ee
$r_{0}=5.3$ {\AA} is the (angle averaged) 
radius of the electron distribution in the molecule based on the 
bulk density of 1.64 g/cm$^3$ \cite{KolomenskiiASS1995} and a FCC 
packing ratio of 0.74 \cite{AshcroftSSP}.
The level densities can be approximated as
\be
\rho^{(q+1)}(E-\veps) \approx \rho^{(q+1)}(E) \e^{-\veps/T(E)}.
\ee
At a given photon energy the energies of the emitting ions are 
\be
E_1 =  h\nu - E_{\rm i,1}
\ee
for the first emitted electron, and
\be
E_2 = h\nu - E_{\rm i,1}-E_{\rm i,2} - T(E_1),
\ee
for the second.
$T(E_1)$ is the average value of the energy carried away by the electron during the 
first ionization.
The temperatures are then found from Eq.\ref{caloric}, which can be used 
for both charge states.
Denoting these temperatures by $T_1$ and $T_2$ the spectra become
\be
\label{theospec}
P(\veps)  \propto 
\frac{\(\veps + 3.0~ {\rm eV}\)\e^{-\veps/T_1}}{T_1\(T_1+ 3.0~ {\rm eV}\)}  
+  
\frac{\(\veps + 6.0 ~ {\rm eV}\)\e^{-\veps/T_2}}{T_2\(T_2 + 6.0 ~ {\rm eV}\)}.
\ee
Given the fairly high temperatures, the term proportional to $\veps$
is needed here.

\begin{figure}
\centering
\vspace{-0.7cm}
\includegraphics[width=0.5\textwidth,angle=0]{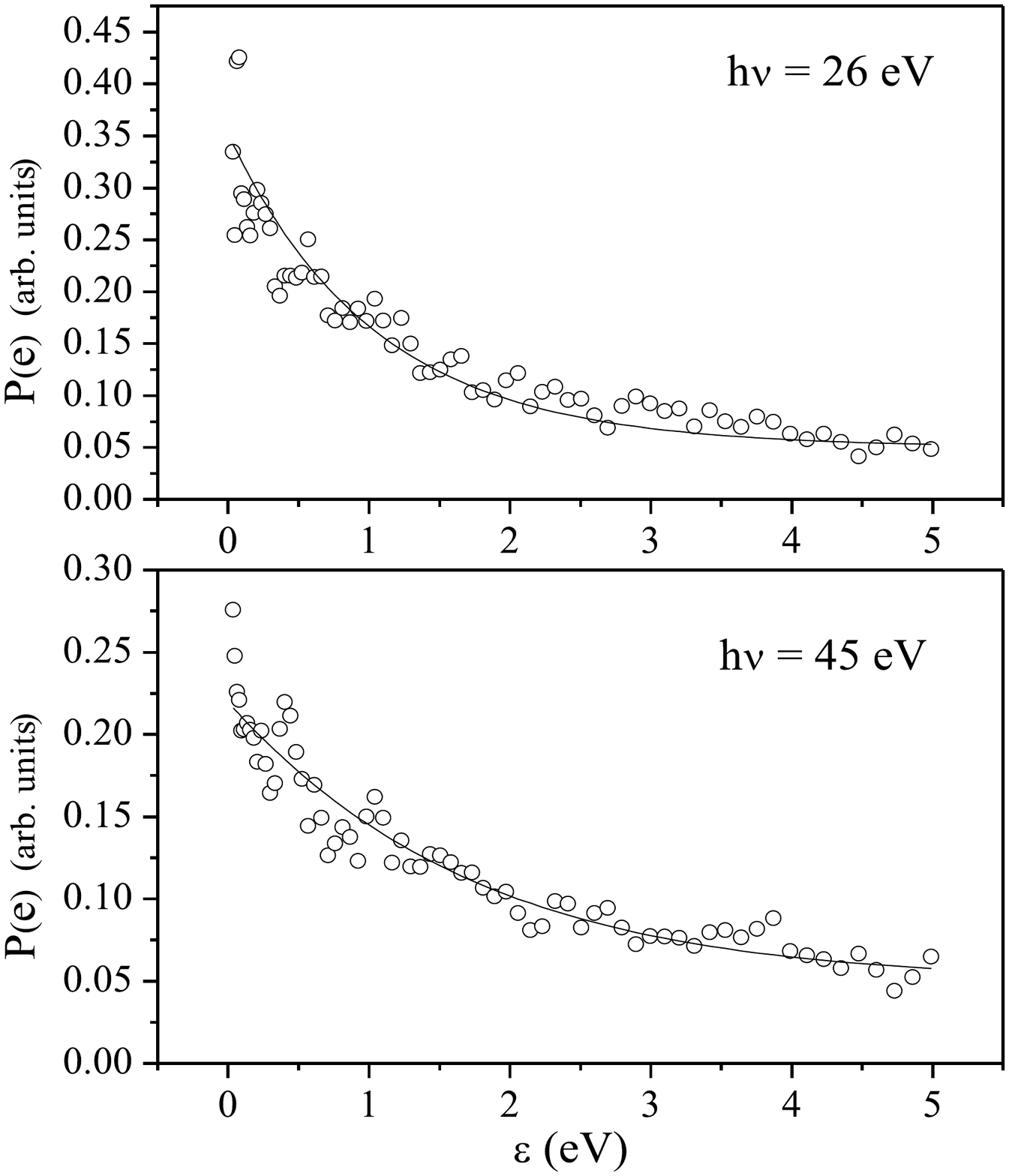}
\vspace{-2.2cm}
\caption{\label{Spectra} 
The low energy parts of the experimental doubly ionized spectra (open circles) 
and the predicted bi-exponential decay in Eq.\ref{theospec}
with the temperatures of  1.053, 0.753 eV for the $h\nu=26$ eV curve,
and 1.391, 1.174 eV for the $h\nu=45$ eV curve, 
given by Eq.\ref{caloric} and calculated with the energies as described
in the main text.}
\end{figure}

The two temperatures in Eq.\ref{theospec} are theoretically rather similar in the 
photon energy range 26 eV to 45 eV.
The values of $T_1$ and $T_2$ for $h\nu=26$ eV, for example, which gives rise to 
the largest difference, are 1.05 eV and 0.75 eV.
For $h\nu=45$ eV they are 1.39 eV and 1.17 eV. 
This makes a direct fit uncertain. 
A compounding complication for a fit is that the spectra are found to contain a small and 
broad background. 
Instead, the theoretical curves are plotted with a constant offset of 0.05.
These are shown in Fig. \ref{Spectra} for the lowest and highest photon energies 
for spectra of doubly ionized molecules. 
The quality of the prediction for the spectra of the four photon energies not shown 
is very similar. 

With the expression for the rate constant we check the consistency of the 
hot electron picture by a comparison of the theory value $\alpha'$ with the 
experimental value $\alpha$ on the right hand side of Eq. \ref{Ehnu2appear}.
The rate constant to choose is the one corresponding to the time constant of
dissipation of the electronic excitation energy into vibrational motion, Eq.\ref{k-limit};
$k(E) = 1/240 \, {\rm fs}$. 
This gives the energy 13 eV. 
Using this for the first ionization gives the corresponding singly ionized version of
Eq. \ref{Ehnu2appear}:
\be
\frac{\alpha}{2 (\ln \omega_{\rm e} \tau)^2} = \frac{13 {\rm eV}}{E_{\rm i,1}} 
= 0.10 ~{\rm eV}^{-1}, 
\ee
which is in reasonable agreement with the experimental value of 0.13 eV$^{-1}$.
We take this as a confirmation of the values used.
It should be noted that clearly this only confirms the product of the electron emission
frequency factor and the value of $\tau$ and not the values of the two quantities 
separately.

Finally we will address the question of the initial excitation of the molecule.
One of the challenges still facing the description of the phenomenon is an 
explanation of the mechanism of the initial excitation.
A part of this question, which will be susceptible to future experiments and
that has obvious implications for the kinetic energies of the emitted electrons,
is the branching ratio between direct and hot electron ionization.
A full quantum mechanical description of the dynamics of the process is 
beyond the scope of this publication, but we will suggest a possible mechanism 
which will convert a single-particle excitation into multi-electron excitation and 
hence provide the initial energy dissipation needed to produce the electron 
spectra seen in this work.

The suggested description builds on the single-electron picture. 
In the initial reaction, the photon is absorbed by a single electron which is 
promoted to a vacuum state, converting all energy in excess of the binding energy, 
$E_{\rm b}$, to kinetic energy according to the standard relation
\be
E_{\rm k} = h\nu - E_{\rm b}.
\ee
After this, the electron starts to move across the fullerene.
The time it takes for this crossing is given classically by
\be
t_{\rm c} \sim \frac{2r_{0}}{\sqrt{2(h\nu - E_{\rm b})/m}}.
\ee
During this motion the remaining valence electrons will be exposed to the 
electric field of the excited electron.
This will excite the surface plasmon resonance with some probability, which 
will depend on the speed of the emitted electron.
Setting $t_{\rm c}$ to half the period of a resonance, the departing electron 
will then be in resonance with an excitation with a quantum energy of
\be
\label{ResonanceCondition}
\hbar \omega = \hbar \pi 
\sqrt{2(h\nu - E_{\rm b})/m} /2 r_{0}.
\ee
A kinetic energy of 50 eV, for example, will give the value of 8.6 eV for the right hand 
side, and be optimal for exciting an oscillation around that energy. 
This energy is on the order of the peak energy of the surface plasmon resonance, which 
is located with a centroid energy of $\hbar \omega \approx 20$ eV and, significantly, 
with a width of similar magnitude \cite{HertelPRL1992}.
Electron energy loss spectroscopy shows a
strong absorption of collective nature from 5 eV electron energy and up
\cite{SohmenZPB1992}, similar to the optical cross section. 
The attenuation length was given for C$_{60}$ films in \cite{LiJESRP2006},
for a single energy.
The value compares well with values from intercalated fullerite samples and the
pure fullerite attenuation length at the energies relevant here can be taken with 
some confidence to be around the size of the molecule. 
There is little reason to believe that the value for C$_{70}$ is significantly different,
given the similar spectra in \cite{SohmenZPB1992}. 
These experimental indications suggest that excitation by the prescribed mechanism 
is indeed likely to occur.

Leaving aside the precise value of the matrix element for exciting the plasmon
resonance, it is also clear that at least the time scales match semiclassically.
Moreover, as this resonance is a collective motion of a large number of electrons, 
with the number reflected in the large oscillator strength, a coupling to it  
will deposit the kinetic energy into a large number of valence 
electrons, facilitating the dissipation into incoherent energy which is the 
hallmarks of the hot electron phase.

The mechanism suggested here has some support in the ionization of metal clusters.
In Ref.\cite{WongAPB2001} ionization yields of alkali metal clusters are reported.
The data show reductions in ionization yields above the surface plasmon resonance.
This is discussed qualitatively in terms of a mechanism that couple photo-electrons 
and the plasmon, similar to the one predicted here. 
In particular, it will impact measured ionization cross sections, such as those 
reported in Ref. \cite{HertelPRL1992}, although 
for those measurements the corrections will mainly occur at the high energy side
of the peak value.
An experimental signature of the effect is a reduced direct ionization efficiency 
and an increased amount of hot electron ionization in the energy region where the 
kinetic energy in the initial stage is conducive to excitation of the resonance, 
i.e. fulfills Eq.\ref{ResonanceCondition}.
This is effectively a shadow of the plasmon resonance.
This Resonance Ionization Shadowing must be expected to be present in other 
clusters or molecules that have large oscillator strength resonances.
The precise parameters of the effect, such as the branching ratio of direct ionization
to hot electron formation, will depend on the centroid energy, its width and to 
some extent also on its oscillator strength. 
The molecular geometry may likewise determine the initial coupling to the 
resonance.

\section{Conclusion and outlook}

We have measured the single photon hot electron ionization of 
C$_{70}$. 
It shows the same main features as the process for C$_{60}$, albeit with 
a somewhat stronger intensity of the doubly ionized species compared to fragmentation.
The measurements thus demonstrate that the mechanism is not restricted to 
a single fullerene.
The mechanism by which the molecules absorb a photon with energy above the 
ionization energy and equilibrates it is not yet established. 
In this work we have suggested a mechanism involving excitation of the 
surface plasmon by a departing electron.
This mechanism should be fairly general.
If correct, it will give a suppression of the ionization as a function of  photon 
energy in a wide energy region, usually above the plasmon centroid.
Part of the suppression will be compensated by the enhanced hot electron emission.
The suggested mechanism is not a direct excitation of the plasmon and 
explains that the onset of the hot electron emission appears above its centroid 
energy, as already seen for C$_{60}$. 
Hence, contrary to previous statements, the surface plasmon is relevant after all, 
albeit only indirectly.

\section*{acknowledgement}

KH acknowledges support from NSFC with grant No. 12047501, from 
the 111 Project of the Ministry of Science and Technology of People's Republic of 
China' under grant no. B20063, and Y. Gong for advice on the presentation.
P.F. acknowledges the Research Foundation Flanders (FWO) for a Senior 
postdoctoral grant. 
The computational resources and services used in this work were provided by 
the VSC (Flemish Supercomputer Center), funded by the FWO and the Flemish 
Government.
We acknowledge Elettra Sincrotrone Trieste for providing access to its synchrotron 
radiation facilities. 
The authors also acknowledge the open access contribution of the Research 
Infrastructure (RI) Elettra.
Comments from V. Kresin are gratefully acknowledged.


\begin{thebibliography}{43}
\expandafter\ifx\csname natexlab\endcsname\relax\def\natexlab#1{#1}\fi
\expandafter\ifx\csname bibnamefont\endcsname\relax
  \def\bibnamefont#1{#1}\fi
\expandafter\ifx\csname bibfnamefont\endcsname\relax
  \def\bibfnamefont#1{#1}\fi
\expandafter\ifx\csname citenamefont\endcsname\relax
  \def\citenamefont#1{#1}\fi
\expandafter\ifx\csname url\endcsname\relax
  \def\url#1{\texttt{#1}}\fi
\expandafter\ifx\csname urlprefix\endcsname\relax\def\urlprefix{URL }\fi
\providecommand{\bibinfo}[2]{#2}
\providecommand{\eprint}[2][]{\url{#2}}

\bibitem[{\citenamefont{Anisimov et~al.}(1974)\citenamefont{Anisimov,
  Kapeliovich, and Perel'man}}]{AnisimovJETP1974}
\bibinfo{author}{\bibfnamefont{S.~I.} \bibnamefont{Anisimov}},
  \bibinfo{author}{\bibfnamefont{B.~L.} \bibnamefont{Kapeliovich}},
  \bibnamefont{and} \bibinfo{author}{\bibfnamefont{T.~L.}
  \bibnamefont{Perel'man}}, \bibinfo{journal}{Sov. Phys. JETP}
  \textbf{\bibinfo{volume}{39}}, \bibinfo{pages}{375} (\bibinfo{year}{1974}).

\bibitem[{\citenamefont{Weber et~al.}(1998)\citenamefont{Weber, Hansen, Ruf,
  and Hotop}}]{WeberCP1998}
\bibinfo{author}{\bibfnamefont{J.~M.} \bibnamefont{Weber}},
  \bibinfo{author}{\bibfnamefont{K.}~\bibnamefont{Hansen}},
  \bibinfo{author}{\bibfnamefont{M.~W.} \bibnamefont{Ruf}}, \bibnamefont{and}
  \bibinfo{author}{\bibfnamefont{H.}~\bibnamefont{Hotop}},
  \bibinfo{journal}{Chem. Phys.} \textbf{\bibinfo{volume}{239}},
  \bibinfo{pages}{271} (\bibinfo{year}{1998}).

\bibitem[{\citenamefont{Campbell et~al.}(2000)\citenamefont{Campbell, Hansen,
  Hoffmann, Korn, Tchaplyguine, Wittmann, and Hertel}}]{CampbellPRL2000}
\bibinfo{author}{\bibfnamefont{E.~E.~B.} \bibnamefont{Campbell}},
  \bibinfo{author}{\bibfnamefont{K.}~\bibnamefont{Hansen}},
  \bibinfo{author}{\bibfnamefont{K.}~\bibnamefont{Hoffmann}},
  \bibinfo{author}{\bibfnamefont{G.}~\bibnamefont{Korn}},
  \bibinfo{author}{\bibfnamefont{M.}~\bibnamefont{Tchaplyguine}},
  \bibinfo{author}{\bibfnamefont{M.}~\bibnamefont{Wittmann}}, \bibnamefont{and}
  \bibinfo{author}{\bibfnamefont{I.~V.} \bibnamefont{Hertel}},
  \bibinfo{journal}{Phys. Rev. Lett.} \textbf{\bibinfo{volume}{84}},
  \bibinfo{pages}{2128} (\bibinfo{year}{2000}).

\bibitem[{\citenamefont{Schlipper et~al.}(2001)\citenamefont{Schlipper, Kusche,
  v.~Issendorff, and Haberland}}]{SchlipperAPA2001}
\bibinfo{author}{\bibfnamefont{R.}~\bibnamefont{Schlipper}},
  \bibinfo{author}{\bibfnamefont{R.}~\bibnamefont{Kusche}},
  \bibinfo{author}{\bibfnamefont{B.}~\bibnamefont{v.~Issendorff}},
  \bibnamefont{and}
  \bibinfo{author}{\bibfnamefont{H.}~\bibnamefont{Haberland}},
  \bibinfo{journal}{Appl. Phys. A} \textbf{\bibinfo{volume}{72}},
  \bibinfo{pages}{255} (\bibinfo{year}{2001}).

\bibitem[{\citenamefont{Maier et~al.}(2006{\natexlab{a}})\citenamefont{Maier,
  Sch{\"a}tzel, Wrigge, Astruc~Hoffmann, Didier, and
  v.~Issendorff}}]{MaierIJMS2006}
\bibinfo{author}{\bibfnamefont{M.}~\bibnamefont{Maier}},
  \bibinfo{author}{\bibfnamefont{M.}~\bibnamefont{Sch{\"a}tzel}},
  \bibinfo{author}{\bibfnamefont{G.}~\bibnamefont{Wrigge}},
  \bibinfo{author}{\bibfnamefont{M.}~\bibnamefont{Astruc~Hoffmann}},
  \bibinfo{author}{\bibfnamefont{P.}~\bibnamefont{Didier}}, \bibnamefont{and}
  \bibinfo{author}{\bibfnamefont{B.}~\bibnamefont{v.~Issendorff}},
  \bibinfo{journal}{Int. J. Mass Spectrom.} \textbf{\bibinfo{volume}{252}},
  \bibinfo{pages}{157} (\bibinfo{year}{2006}{\natexlab{a}}).

\bibitem[{\citenamefont{Maier et~al.}(2006{\natexlab{b}})\citenamefont{Maier,
  Wrigge, Astruc~Hoffmann, Didier, and v.~Issendorff}}]{MaierPRL2006}
\bibinfo{author}{\bibfnamefont{M.}~\bibnamefont{Maier}},
  \bibinfo{author}{\bibfnamefont{G.}~\bibnamefont{Wrigge}},
  \bibinfo{author}{\bibfnamefont{M.}~\bibnamefont{Astruc~Hoffmann}},
  \bibinfo{author}{\bibfnamefont{P.}~\bibnamefont{Didier}}, \bibnamefont{and}
  \bibinfo{author}{\bibfnamefont{B.}~\bibnamefont{v.~Issendorff}},
  \bibinfo{journal}{Phys. Rev. Lett.} \textbf{\bibinfo{volume}{96}},
  \bibinfo{pages}{117405} (\bibinfo{year}{2006}{\natexlab{b}}).

\bibitem[{\citenamefont{Kjellberg
  et~al.}(2010{\natexlab{a}})\citenamefont{Kjellberg, Johansson, Jonsson,
  Bulgakov, Bordas, Campbell, and Hansen}}]{KjellbergPRA2010}
\bibinfo{author}{\bibfnamefont{M.}~\bibnamefont{Kjellberg}},
  \bibinfo{author}{\bibfnamefont{O.}~\bibnamefont{Johansson}},
  \bibinfo{author}{\bibfnamefont{F.}~\bibnamefont{Jonsson}},
  \bibinfo{author}{\bibfnamefont{A.~V.} \bibnamefont{Bulgakov}},
  \bibinfo{author}{\bibfnamefont{C.}~\bibnamefont{Bordas}},
  \bibinfo{author}{\bibfnamefont{E.~E.~B.} \bibnamefont{Campbell}},
  \bibnamefont{and} \bibinfo{author}{\bibfnamefont{K.}~\bibnamefont{Hansen}},
  \bibinfo{journal}{Phys. Rev. A} \textbf{\bibinfo{volume}{81}},
  \bibinfo{pages}{023202} (\bibinfo{year}{2010}{\natexlab{a}}).

\bibitem[{\citenamefont{Kjellberg
  et~al.}(2010{\natexlab{b}})\citenamefont{Kjellberg, Bulgakov, Goto,
  johansson, and Hansen}}]{KjellbergJCP2010}
\bibinfo{author}{\bibfnamefont{M.}~\bibnamefont{Kjellberg}},
  \bibinfo{author}{\bibfnamefont{A.~V.} \bibnamefont{Bulgakov}},
  \bibinfo{author}{\bibfnamefont{M.}~\bibnamefont{Goto}},
  \bibinfo{author}{\bibfnamefont{O.}~\bibnamefont{johansson}},
  \bibnamefont{and} \bibinfo{author}{\bibfnamefont{K.}~\bibnamefont{Hansen}},
  \bibinfo{journal}{J. Chem. Phys.} \textbf{\bibinfo{volume}{133}},
  \bibinfo{pages}{074308} (\bibinfo{year}{2010}{\natexlab{b}}).

\bibitem[{\citenamefont{Flambaum et~al.}(2002)\citenamefont{Flambaum,
  Gribakina, Gribakin, and Harabati}}]{FlambaumPRA2002}
\bibinfo{author}{\bibfnamefont{V.~V.} \bibnamefont{Flambaum}},
  \bibinfo{author}{\bibfnamefont{A.~A.} \bibnamefont{Gribakina}},
  \bibinfo{author}{\bibfnamefont{G.~F.} \bibnamefont{Gribakin}},
  \bibnamefont{and} \bibinfo{author}{\bibfnamefont{C.}~\bibnamefont{Harabati}},
  \bibinfo{journal}{Phys. Rev. A} \textbf{\bibinfo{volume}{66}},
  \bibinfo{pages}{012713} (\bibinfo{year}{2002}).

\bibitem[{\citenamefont{Gribakin and Sahoo}(2003)}]{GribakinJPB2003}
\bibinfo{author}{\bibfnamefont{G.}~\bibnamefont{Gribakin}} \bibnamefont{and}
  \bibinfo{author}{\bibfnamefont{S.}~\bibnamefont{Sahoo}},
  \bibinfo{journal}{Journal of Physics B: Atomic Molecular and Optical Physics}
  \textbf{\bibinfo{volume}{36}}, \bibinfo{pages}{3349} (\bibinfo{year}{2003}).

\bibitem[{\citenamefont{Heraud et~al.}(2021)\citenamefont{Heraud, Vincendon,
  Reinhard, Dinh, and Suraud}}]{HeraudEPJD2021}
\bibinfo{author}{\bibfnamefont{J.}~\bibnamefont{Heraud}},
  \bibinfo{author}{\bibfnamefont{M.}~\bibnamefont{Vincendon}},
  \bibinfo{author}{\bibfnamefont{P.-G.} \bibnamefont{Reinhard}},
  \bibinfo{author}{\bibfnamefont{P.~M.} \bibnamefont{Dinh}}, \bibnamefont{and}
  \bibinfo{author}{\bibfnamefont{E.}~\bibnamefont{Suraud}},
  \bibinfo{journal}{Eur. Phys. J. D} \textbf{\bibinfo{volume}{75}}
  (\bibinfo{year}{2021}).

\bibitem[{\citenamefont{Campbell et~al.}(2006)\citenamefont{Campbell, Hansen,
  Hed\'{e}n, Kjellberg, and A.V.Bulgakov}}]{CampbellPPS2006}
\bibinfo{author}{\bibfnamefont{E.~E.~B.} \bibnamefont{Campbell}},
  \bibinfo{author}{\bibfnamefont{K.}~\bibnamefont{Hansen}},
  \bibinfo{author}{\bibfnamefont{M.}~\bibnamefont{Hed\'{e}n}},
  \bibinfo{author}{\bibfnamefont{M.}~\bibnamefont{Kjellberg}},
  \bibnamefont{and} \bibinfo{author}{\bibnamefont{A.V.Bulgakov}},
  \bibinfo{journal}{Photochem. Photobiol. Sci.} \textbf{\bibinfo{volume}{5}},
  \bibinfo{pages}{1183} (\bibinfo{year}{2006}).

\bibitem[{\citenamefont{Hansen et~al.}(2003)\citenamefont{Hansen, Hoffmann, and
  Campbell}}]{HansenJCP2003}
\bibinfo{author}{\bibfnamefont{K.}~\bibnamefont{Hansen}},
  \bibinfo{author}{\bibfnamefont{K.}~\bibnamefont{Hoffmann}}, \bibnamefont{and}
  \bibinfo{author}{\bibfnamefont{E.~E.~B.} \bibnamefont{Campbell}},
  \bibinfo{journal}{J. Chem. Phys.} \textbf{\bibinfo{volume}{119}},
  \bibinfo{pages}{2513} (\bibinfo{year}{2003}).

\bibitem[{\citenamefont{Schettino et~al.}(2001)\citenamefont{Schettino,
  Pagliai, Ciabini, and Cardini}}]{SchettinoJPCA2001}
\bibinfo{author}{\bibfnamefont{V.}~\bibnamefont{Schettino}},
  \bibinfo{author}{\bibfnamefont{M.}~\bibnamefont{Pagliai}},
  \bibinfo{author}{\bibfnamefont{L.}~\bibnamefont{Ciabini}}, \bibnamefont{and}
  \bibinfo{author}{\bibfnamefont{G.}~\bibnamefont{Cardini}},
  \bibinfo{journal}{J. Phys. Chem. A} \textbf{\bibinfo{volume}{105}},
  \bibinfo{pages}{11192} (\bibinfo{year}{2001}).

\bibitem[{\citenamefont{Mongin et~al.}(2019)\citenamefont{Mongin, Maioli,
  Burgin, Langot, Cottancin, D'Addato, Canut, Treguer, Crut, Vall{\'e}e
  et~al.}}]{MonginJPCM2019}
\bibinfo{author}{\bibfnamefont{D.}~\bibnamefont{Mongin}},
  \bibinfo{author}{\bibfnamefont{P.}~\bibnamefont{Maioli}},
  \bibinfo{author}{\bibfnamefont{J.}~\bibnamefont{Burgin}},
  \bibinfo{author}{\bibfnamefont{P.}~\bibnamefont{Langot}},
  \bibinfo{author}{\bibfnamefont{E.}~\bibnamefont{Cottancin}},
  \bibinfo{author}{\bibfnamefont{S.}~\bibnamefont{D'Addato}},
  \bibinfo{author}{\bibfnamefont{B.}~\bibnamefont{Canut}},
  \bibinfo{author}{\bibfnamefont{M.}~\bibnamefont{Treguer}},
  \bibinfo{author}{\bibfnamefont{A.}~\bibnamefont{Crut}},
  \bibinfo{author}{\bibfnamefont{F.}~\bibnamefont{Vall{\'e}e}},
  \bibnamefont{et~al.}, \bibinfo{journal}{J.Phys. Condens. Matter}
  \textbf{\bibinfo{volume}{31}}, \bibinfo{pages}{084001}
  (\bibinfo{year}{2019}).

\bibitem[{\citenamefont{Cong-fang et~al.}(1996)\citenamefont{Cong-fang,
  Xi-cheng, Zong-ju, and Ying-hua}}]{WangCPL1996}
\bibinfo{author}{\bibfnamefont{W.}~\bibnamefont{Cong-fang}},
  \bibinfo{author}{\bibfnamefont{A.}~\bibnamefont{Xi-cheng}},
  \bibinfo{author}{\bibfnamefont{X.}~\bibnamefont{Zong-ju}}, \bibnamefont{and}
  \bibinfo{author}{\bibfnamefont{Z.}~\bibnamefont{Ying-hua}},
  \bibinfo{journal}{Chinese Phys. Lett.} \textbf{\bibinfo{volume}{13}},
  \bibinfo{pages}{668} (\bibinfo{year}{1996}).

\bibitem[{\citenamefont{Hansen et~al.}(2017)\citenamefont{Hansen, Richter,
  Alagia, Stranges, Schio, Sal{\'e}n, Yatsyna, Feifel, and
  Zhaunerchyk}}]{HansenPRL2017}
\bibinfo{author}{\bibfnamefont{K.}~\bibnamefont{Hansen}},
  \bibinfo{author}{\bibfnamefont{R.}~\bibnamefont{Richter}},
  \bibinfo{author}{\bibfnamefont{M.}~\bibnamefont{Alagia}},
  \bibinfo{author}{\bibfnamefont{S.}~\bibnamefont{Stranges}},
  \bibinfo{author}{\bibfnamefont{L.}~\bibnamefont{Schio}},
  \bibinfo{author}{\bibfnamefont{P.}~\bibnamefont{Sal{\'e}n}},
  \bibinfo{author}{\bibfnamefont{V.}~\bibnamefont{Yatsyna}},
  \bibinfo{author}{\bibfnamefont{R.}~\bibnamefont{Feifel}}, \bibnamefont{and}
  \bibinfo{author}{\bibfnamefont{V.}~\bibnamefont{Zhaunerchyk}},
  \bibinfo{journal}{Phys. Rev. Lett.} \textbf{\bibinfo{volume}{118}},
  \bibinfo{pages}{103001} (\bibinfo{year}{2017}).

\bibitem[{\citenamefont{Herbst}(2005)}]{Herbst_2005}
\bibinfo{author}{\bibfnamefont{E.}~\bibnamefont{Herbst}},
  \bibinfo{journal}{Journal of Physics: Conference Series}
  \textbf{\bibinfo{volume}{4}}, \bibinfo{pages}{17} (\bibinfo{year}{2005}).

\bibitem[{\citenamefont{Andersen et~al.}(2002)\citenamefont{Andersen, Bonderup,
  and Hansen}}]{AndersenJPB2002}
\bibinfo{author}{\bibfnamefont{J.~U.} \bibnamefont{Andersen}},
  \bibinfo{author}{\bibfnamefont{E.}~\bibnamefont{Bonderup}}, \bibnamefont{and}
  \bibinfo{author}{\bibfnamefont{K.}~\bibnamefont{Hansen}},
  \bibinfo{journal}{J. Phys. B} \textbf{\bibinfo{volume}{35}},
  \bibinfo{pages}{R1} (\bibinfo{year}{2002}).

\bibitem[{\citenamefont{Andersen et~al.}(2001)\citenamefont{Andersen, Bonderup,
  and Hansen}}]{AndersenJCP2001}
\bibinfo{author}{\bibfnamefont{J.~U.} \bibnamefont{Andersen}},
  \bibinfo{author}{\bibfnamefont{E.}~\bibnamefont{Bonderup}}, \bibnamefont{and}
  \bibinfo{author}{\bibfnamefont{K.}~\bibnamefont{Hansen}},
  \bibinfo{journal}{J. Chem. Phys.} \textbf{\bibinfo{volume}{114}},
  \bibinfo{pages}{6518} (\bibinfo{year}{2001}).

\bibitem[{\citenamefont{Hansen}(2018)}]{book}
\bibinfo{author}{\bibfnamefont{K.}~\bibnamefont{Hansen}},
  \emph{\bibinfo{title}{{Statistical Physics of Nanoparticles in the Gas
  Phase}{}}}, vol.~\bibinfo{volume}{73} of \emph{\bibinfo{series}{Springer
  Series on Atomic, Optical, and Plasma Physics}}
  (\bibinfo{publisher}{Springer}, \bibinfo{address}{Dordrecht},
  \bibinfo{year}{2018}).

\bibitem[{\citenamefont{L{\'e}pine and Bordas}(2004)}]{LepinePRA2004}
\bibinfo{author}{\bibfnamefont{F.}~\bibnamefont{L{\'e}pine}} \bibnamefont{and}
  \bibinfo{author}{\bibfnamefont{C.}~\bibnamefont{Bordas}},
  \bibinfo{journal}{Phys. Rev. A} \textbf{\bibinfo{volume}{69}},
  \bibinfo{pages}{053201} (\bibinfo{year}{2004}).

\bibitem[{\citenamefont{Walder and Echt}(1992)}]{WalderIJMPB1992}
\bibinfo{author}{\bibfnamefont{G.}~\bibnamefont{Walder}} \bibnamefont{and}
  \bibinfo{author}{\bibfnamefont{O.}~\bibnamefont{Echt}},
  \bibinfo{journal}{Int. J. Modern Physics B} \textbf{\bibinfo{volume}{6}},
  \bibinfo{pages}{3881} (\bibinfo{year}{1992}).

\bibitem[{\citenamefont{Andersson et~al.}(2019)\citenamefont{Andersson,
  Zagorodskikh, Roos, Talaee, Squibb, Koulentianos, Wallner, Zhaunerchyk,
  Singh, Eland et~al.}}]{AnderssonSR2019}
\bibinfo{author}{\bibfnamefont{J.}~\bibnamefont{Andersson}},
  \bibinfo{author}{\bibfnamefont{S.}~\bibnamefont{Zagorodskikh}},
  \bibinfo{author}{\bibfnamefont{A.~H.} \bibnamefont{Roos}},
  \bibinfo{author}{\bibfnamefont{O.}~\bibnamefont{Talaee}},
  \bibinfo{author}{\bibfnamefont{R.~J.} \bibnamefont{Squibb}},
  \bibinfo{author}{\bibfnamefont{D.}~\bibnamefont{Koulentianos}},
  \bibinfo{author}{\bibfnamefont{M.}~\bibnamefont{Wallner}},
  \bibinfo{author}{\bibfnamefont{V.}~\bibnamefont{Zhaunerchyk}},
  \bibinfo{author}{\bibfnamefont{R.}~\bibnamefont{Singh}},
  \bibinfo{author}{\bibfnamefont{J.~H.~D.} \bibnamefont{Eland}},
  \bibnamefont{et~al.}, \bibinfo{journal}{Sci. Rep.}
  \textbf{\bibinfo{volume}{9}}, \bibinfo{pages}{17883} (\bibinfo{year}{2019}).

\bibitem[{\citenamefont{Lassesson et~al.}(2005)\citenamefont{Lassesson, Hansen,
  J\"{o}nsson, Gromov, Campbell, Boyle, Pop, Schulz, Hertel, Taninaka
  et~al.}}]{LassessonEPJD2005}
\bibinfo{author}{\bibfnamefont{A.}~\bibnamefont{Lassesson}},
  \bibinfo{author}{\bibfnamefont{K.}~\bibnamefont{Hansen}},
  \bibinfo{author}{\bibfnamefont{M.}~\bibnamefont{J\"{o}nsson}},
  \bibinfo{author}{\bibfnamefont{A.}~\bibnamefont{Gromov}},
  \bibinfo{author}{\bibfnamefont{E.}~\bibnamefont{Campbell}},
  \bibinfo{author}{\bibfnamefont{M.}~\bibnamefont{Boyle}},
  \bibinfo{author}{\bibfnamefont{D.}~\bibnamefont{Pop}},
  \bibinfo{author}{\bibfnamefont{C.~P.} \bibnamefont{Schulz}},
  \bibinfo{author}{\bibfnamefont{I.~V.} \bibnamefont{Hertel}},
  \bibinfo{author}{\bibfnamefont{A.}~\bibnamefont{Taninaka}},
  \bibnamefont{et~al.}, \bibinfo{journal}{Eur. Phys. J. D}
  \textbf{\bibinfo{volume}{34}}, \bibinfo{pages}{205} (\bibinfo{year}{2005}).

\bibitem[{\citenamefont{Hrodmarsson et~al.}(2020)\citenamefont{Hrodmarsson,
  Garcia, Linnartz, and Nahom}}]{HrodmarssonPCCP2020}
\bibinfo{author}{\bibfnamefont{H.~R.} \bibnamefont{Hrodmarsson}},
  \bibinfo{author}{\bibfnamefont{G.~A.} \bibnamefont{Garcia}},
  \bibinfo{author}{\bibfnamefont{H.}~\bibnamefont{Linnartz}}, \bibnamefont{and}
  \bibinfo{author}{\bibfnamefont{L.}~\bibnamefont{Nahom}},
  \bibinfo{journal}{Phys. Chem. Chem. Phys.} \textbf{\bibinfo{volume}{22}},
  \bibinfo{pages}{13880} (\bibinfo{year}{2020}).

\bibitem[{\citenamefont{Blyth et~al.}(1999)\citenamefont{Blyth, Delaunay,
  Zitnik, Krempasky, Krempaska, Slezak, Prince, Richter, Vondracek, Camilloni
  et~al.}}]{BlythESRP1999}
\bibinfo{author}{\bibfnamefont{R.~R.} \bibnamefont{Blyth}},
  \bibinfo{author}{\bibfnamefont{R.}~\bibnamefont{Delaunay}},
  \bibinfo{author}{\bibfnamefont{M.}~\bibnamefont{Zitnik}},
  \bibinfo{author}{\bibfnamefont{J.}~\bibnamefont{Krempasky}},
  \bibinfo{author}{\bibfnamefont{R.}~\bibnamefont{Krempaska}},
  \bibinfo{author}{\bibfnamefont{J.}~\bibnamefont{Slezak}},
  \bibinfo{author}{\bibfnamefont{K.~C.} \bibnamefont{Prince}},
  \bibinfo{author}{\bibfnamefont{R.}~\bibnamefont{Richter}},
  \bibinfo{author}{\bibfnamefont{M.}~\bibnamefont{Vondracek}},
  \bibinfo{author}{\bibfnamefont{R.}~\bibnamefont{Camilloni}},
  \bibnamefont{et~al.}, \bibinfo{journal}{J. Electron Spectrosc. Relat.
  Phenom.} \textbf{\bibinfo{volume}{101-103}}, \bibinfo{pages}{959}
  (\bibinfo{year}{1999}).

\bibitem[{\citenamefont{O'Keeffe et~al.}(2011)\citenamefont{O'Keeffe,
  Bolognesi, Coreno, Moise, Richter, Cautero, Stebel, Sergo, Pravica,
  Ovcharenko et~al.}}]{OkeeffeRSI2011}
\bibinfo{author}{\bibfnamefont{P.}~\bibnamefont{O'Keeffe}},
  \bibinfo{author}{\bibfnamefont{P.}~\bibnamefont{Bolognesi}},
  \bibinfo{author}{\bibfnamefont{M.}~\bibnamefont{Coreno}},
  \bibinfo{author}{\bibfnamefont{A.}~\bibnamefont{Moise}},
  \bibinfo{author}{\bibfnamefont{R.}~\bibnamefont{Richter}},
  \bibinfo{author}{\bibfnamefont{G.}~\bibnamefont{Cautero}},
  \bibinfo{author}{\bibfnamefont{L.}~\bibnamefont{Stebel}},
  \bibinfo{author}{\bibfnamefont{R.}~\bibnamefont{Sergo}},
  \bibinfo{author}{\bibfnamefont{L.}~\bibnamefont{Pravica}},
  \bibinfo{author}{\bibfnamefont{Y.}~\bibnamefont{Ovcharenko}},
  \bibnamefont{et~al.}, \bibinfo{journal}{Rev. Sci. Instrum.}
  \textbf{\bibinfo{volume}{82}}, \bibinfo{pages}{033109}
  (\bibinfo{year}{2011}).

\bibitem[{\citenamefont{Dick}(2014)}]{DickPCCP2014}
\bibinfo{author}{\bibfnamefont{B.}~\bibnamefont{Dick}}, \bibinfo{journal}{Phys.
  Chem. Chem. Phys.} \textbf{\bibinfo{volume}{16}}, \bibinfo{pages}{570}
  (\bibinfo{year}{2014}).

\bibitem[{\citenamefont{Boltalina et~al.}(2000)\citenamefont{Boltalina, Ioffe,
  Sidorov, Seifert, , and Vietze}}]{BoltalinaJACS2000}
\bibinfo{author}{\bibfnamefont{O.~V.} \bibnamefont{Boltalina}},
  \bibinfo{author}{\bibfnamefont{I.~N.} \bibnamefont{Ioffe}},
  \bibinfo{author}{\bibfnamefont{L.~N.} \bibnamefont{Sidorov}},
  \bibinfo{author}{\bibfnamefont{G.}~\bibnamefont{Seifert}}, ,
  \bibnamefont{and} \bibinfo{author}{\bibfnamefont{K.}~\bibnamefont{Vietze}},
  \bibinfo{journal}{J. Am. Chem. Soc.} \textbf{\bibinfo{volume}{122}},
  \bibinfo{pages}{9745} (\bibinfo{year}{2000}).

\bibitem[{\citenamefont{Steger et~al.}(1992)\citenamefont{Steger, de~Vries,
  Kamke, Kamke, and Drewello}}]{StegerCPL1992}
\bibinfo{author}{\bibfnamefont{H.}~\bibnamefont{Steger}},
  \bibinfo{author}{\bibfnamefont{J.}~\bibnamefont{de~Vries}},
  \bibinfo{author}{\bibfnamefont{B.}~\bibnamefont{Kamke}},
  \bibinfo{author}{\bibfnamefont{W.}~\bibnamefont{Kamke}}, \bibnamefont{and}
  \bibinfo{author}{\bibfnamefont{T.}~\bibnamefont{Drewello}},
  \bibinfo{journal}{Chem. Phys. Lett.} \textbf{\bibinfo{volume}{194}}
  (\bibinfo{year}{1992}).

\bibitem[{\citenamefont{Neese et~al.}(2020)\citenamefont{Neese, Wennmohs,
  Becker, and Riplinger}}]{neese2020orca}
\bibinfo{author}{\bibfnamefont{F.}~\bibnamefont{Neese}},
  \bibinfo{author}{\bibfnamefont{F.}~\bibnamefont{Wennmohs}},
  \bibinfo{author}{\bibfnamefont{U.}~\bibnamefont{Becker}}, \bibnamefont{and}
  \bibinfo{author}{\bibfnamefont{C.}~\bibnamefont{Riplinger}},
  \bibinfo{journal}{J. Chem. Phys.} \textbf{\bibinfo{volume}{152}},
  \bibinfo{pages}{224108} (\bibinfo{year}{2020}).

\bibitem[{\citenamefont{Hammer et~al.}(1999)\citenamefont{Hammer, Hansen, and
  N{\o}rskov}}]{hammer1999improved}
\bibinfo{author}{\bibfnamefont{B.}~\bibnamefont{Hammer}},
  \bibinfo{author}{\bibfnamefont{L.~B.} \bibnamefont{Hansen}},
  \bibnamefont{and} \bibinfo{author}{\bibfnamefont{J.~K.}
  \bibnamefont{N{\o}rskov}}, \bibinfo{journal}{Phys. Rev. B}
  \textbf{\bibinfo{volume}{59}}, \bibinfo{pages}{7413} (\bibinfo{year}{1999}).

\bibitem[{\citenamefont{Weigend and Ahlrichs}(2005)}]{weigend2005balanced}
\bibinfo{author}{\bibfnamefont{F.}~\bibnamefont{Weigend}} \bibnamefont{and}
  \bibinfo{author}{\bibfnamefont{R.}~\bibnamefont{Ahlrichs}},
  \bibinfo{journal}{Phys. Chem. Chem. Phys.} \textbf{\bibinfo{volume}{7}},
  \bibinfo{pages}{3297} (\bibinfo{year}{2005}).

\bibitem[{\citenamefont{Grimme et~al.}(2011)\citenamefont{Grimme, Ehrlich, and
  Goerigk}}]{grimme2011effect}
\bibinfo{author}{\bibfnamefont{S.}~\bibnamefont{Grimme}},
  \bibinfo{author}{\bibfnamefont{S.}~\bibnamefont{Ehrlich}}, \bibnamefont{and}
  \bibinfo{author}{\bibfnamefont{L.}~\bibnamefont{Goerigk}},
  \bibinfo{journal}{J. Comput.} \textbf{\bibinfo{volume}{32}},
  \bibinfo{pages}{1456} (\bibinfo{year}{2011}).

\bibitem[{\citenamefont{Mitsuke et~al.}(2007)\citenamefont{Mitsuke, Katayanagi,
  Kafle, Huang, Yagi, Prodhan, and Kubozono}}]{MitsukeJPCA2007}
\bibinfo{author}{\bibfnamefont{K.}~\bibnamefont{Mitsuke}},
  \bibinfo{author}{\bibfnamefont{H.}~\bibnamefont{Katayanagi}},
  \bibinfo{author}{\bibfnamefont{B.~P.} \bibnamefont{Kafle}},
  \bibinfo{author}{\bibfnamefont{C.}~\bibnamefont{Huang}},
  \bibinfo{author}{\bibfnamefont{H.}~\bibnamefont{Yagi}},
  \bibinfo{author}{\bibfnamefont{M.~S.~I.} \bibnamefont{Prodhan}},
  \bibnamefont{and} \bibinfo{author}{\bibfnamefont{Y.}~\bibnamefont{Kubozono}},
  \bibinfo{journal}{J. Phys. Chem. A} \textbf{\bibinfo{volume}{111}},
  \bibinfo{pages}{8336} (\bibinfo{year}{2007}).

\bibitem[{\citenamefont{Reitsma et~al.}(2019)\citenamefont{Reitsma, Hummert,
  Dura, Loriot, Vrakking, pine, and Kornilov}}]{ReitsmaJPCA2019}
\bibinfo{author}{\bibfnamefont{G.}~\bibnamefont{Reitsma}},
  \bibinfo{author}{\bibfnamefont{J.}~\bibnamefont{Hummert}},
  \bibinfo{author}{\bibfnamefont{J.}~\bibnamefont{Dura}},
  \bibinfo{author}{\bibfnamefont{V.}~\bibnamefont{Loriot}},
  \bibinfo{author}{\bibfnamefont{M.~J.~J.} \bibnamefont{Vrakking}},
  \bibinfo{author}{\bibfnamefont{F.~L.} \bibnamefont{pine}}, \bibnamefont{and}
  \bibinfo{author}{\bibfnamefont{O.}~\bibnamefont{Kornilov}},
  \bibinfo{journal}{J. Phys. Chem. A} \textbf{\bibinfo{volume}{123}},
  \bibinfo{pages}{3068} (\bibinfo{year}{2019}).

\bibitem[{\citenamefont{Kolomenskii et~al.}(1995)\citenamefont{Kolomenskii,
  Szabadi, and Hess}}]{KolomenskiiASS1995}
\bibinfo{author}{\bibfnamefont{A.~A.} \bibnamefont{Kolomenskii}},
  \bibinfo{author}{\bibfnamefont{M.}~\bibnamefont{Szabadi}}, \bibnamefont{and}
  \bibinfo{author}{\bibfnamefont{P.}~\bibnamefont{Hess}},
  \bibinfo{journal}{App. Surf. Sci.} \textbf{\bibinfo{volume}{86}},
  \bibinfo{pages}{591} (\bibinfo{year}{1995}).

\bibitem[{\citenamefont{Ashcroft and Mermin}(1976)}]{AshcroftSSP}
\bibinfo{author}{\bibfnamefont{N.~W.} \bibnamefont{Ashcroft}} \bibnamefont{and}
  \bibinfo{author}{\bibfnamefont{N.~D.} \bibnamefont{Mermin}},
  \emph{\bibinfo{title}{Solid State Physics}} (\bibinfo{publisher}{Saunder
  College Publishing}, \bibinfo{year}{1976}).

\bibitem[{\citenamefont{Hertel et~al.}(1992)\citenamefont{Hertel, Steger,
  de~Vries, Weisser, Menzel, Kamke, and Kamke}}]{HertelPRL1992}
\bibinfo{author}{\bibfnamefont{I.~V.} \bibnamefont{Hertel}},
  \bibinfo{author}{\bibfnamefont{H.}~\bibnamefont{Steger}},
  \bibinfo{author}{\bibfnamefont{J.}~\bibnamefont{de~Vries}},
  \bibinfo{author}{\bibfnamefont{B.}~\bibnamefont{Weisser}},
  \bibinfo{author}{\bibfnamefont{C.}~\bibnamefont{Menzel}},
  \bibinfo{author}{\bibfnamefont{B.}~\bibnamefont{Kamke}}, \bibnamefont{and}
  \bibinfo{author}{\bibfnamefont{W.}~\bibnamefont{Kamke}},
  \bibinfo{journal}{Phys. Rev. Lett.} \textbf{\bibinfo{volume}{68}}
  (\bibinfo{year}{1992}).

\bibitem[{\citenamefont{Sohmen et~al.}(1992)\citenamefont{Sohmen, Fink, and
  Kr{\"a}tschmer}}]{SohmenZPB1992}
\bibinfo{author}{\bibfnamefont{E.}~\bibnamefont{Sohmen}},
  \bibinfo{author}{\bibfnamefont{J.}~\bibnamefont{Fink}}, \bibnamefont{and}
  \bibinfo{author}{\bibfnamefont{W.}~\bibnamefont{Kr{\"a}tschmer}},
  \bibinfo{journal}{Z. Phys. B} \textbf{\bibinfo{volume}{86}},
  \bibinfo{pages}{87} (\bibinfo{year}{1992}).

\bibitem[{\citenamefont{Li et~al.}(2006)\citenamefont{Li, Wang, and
  Ding}}]{LiJESRP2006}
\bibinfo{author}{\bibfnamefont{H.-N.} \bibnamefont{Li}},
  \bibinfo{author}{\bibfnamefont{X.-X.} \bibnamefont{Wang}}, \bibnamefont{and}
  \bibinfo{author}{\bibfnamefont{W.-F.} \bibnamefont{Ding}},
  \bibinfo{journal}{J. Electron Spectros. Relat. Phenomena}
  \textbf{\bibinfo{volume}{153}}, \bibinfo{pages}{96} (\bibinfo{year}{2006}).

\bibitem[{\citenamefont{Wong et~al.}(2001)\citenamefont{Wong, Kasperovich,
  Tikhonov, and Kresin}}]{WongAPB2001}
\bibinfo{author}{\bibfnamefont{K.}~\bibnamefont{Wong}},
  \bibinfo{author}{\bibfnamefont{V.}~\bibnamefont{Kasperovich}},
  \bibinfo{author}{\bibfnamefont{G.}~\bibnamefont{Tikhonov}}, \bibnamefont{and}
  \bibinfo{author}{\bibfnamefont{V.~V.} \bibnamefont{Kresin}},
  \bibinfo{journal}{Appl. Phys. B} \textbf{\bibinfo{volume}{73}},
  \bibinfo{pages}{407} (\bibinfo{year}{2001}).

\end{thebibliography}

\end{document}